\documentclass[twocolumn,aps,prd,showpacs,nofootinbib,superscriptaddress,preprintnumbers,floatfix,10pt]{revtex4-1}

\usepackage[english]{babel} 
\usepackage[utf8]{inputenc}
\usepackage[T1]{fontenc}
\usepackage{lmodern}

\usepackage{amsmath}
\usepackage{amsfonts}
\usepackage{amssymb}
\usepackage{slashed}
\usepackage{bbm}

\usepackage{graphicx}
\usepackage{color}
\usepackage{multirow}

\usepackage{hyperref}

\newcommand{\I}{\mathrm{i}}

\newcommand{\im}{\mathrm{Im}}
\newcommand{\re}{\mathrm{Re}}

\newcommand{\SU}{\mathrm{SU}}

\newcommand{\rmf}{\mathrm{f}}
\newcommand{\rmh}{\mathrm{h}}

\newcommand{\mh}{m_{\mathrm{h}}}

\newcommand{\mZ}{m_{\mathrm{Z}}}
\newcommand{\mW}{m_{\mathrm{W}}}

\newcommand{\mf}{m_{\mathrm{f}}}
\newcommand{\rmC}{\mathrm{C}}

\newcommand{\rml}{\mathrm{l}}

\newcommand{\phid}{\phi^{\dagger}}
\newcommand{\varphid}{\varphi^{\dagger}}

\newcommand{\loopB}{\boldsymbol\ell_{\mathrm{B}}}
\newcommand{\loopF}{\boldsymbol\ell_{\mathrm{F}}}

\begin{document}

\title{Gauge-invariant description of the Higgs resonance and its phenomenological implications}

\author{Axel Maas}
\email{axel.maas@uni-graz.at}
\affiliation{Institute of Physics, NAWI Graz, University of Graz, Universit\"atsplatz 5, A-8010 Graz, Austria}

\author{Ren\'{e} Sondenheimer}
\email{rene.sondenheimer@uni-graz.at}
\affiliation{Institute of Physics, NAWI Graz, University of Graz, Universit\"atsplatz 5, A-8010 Graz, Austria}

\begin{abstract}
We investigate the phenomenological consequences of a strict gauge-invariant formulation of the Higgs particle. This requires a description of the observable scalar particle in terms of a bound state structure. Although this seems to be at odds with the common treatment of electroweak particle physics at first glance, the properties of the bound state can be described in a perturbative fashion due to the Fr\"ohlich-Morchio-Strocchi (FMS) framework. In particular a relation between the bound-state Higgs and the elementary Higgs field is obtained within $R_{\xi}$ gauges such that the main quantitative properties of the conventional description reappear in leading order of the FMS expansion. Going beyond leading order, we show that the pole structure of the elementary and the bound-state propagator coincide to all orders in a perturbative expansion. However, slight deviations of scattering amplitudes containing off-shell Higgs contributions can be caused by the internal bound state structure. We perform a consistent perturbative treatment to all orders in the FMS expansion to quantify such deviations and demonstrate how gauge-invariant expressions arrange in a natural way at the one-loop level. This also provides a gauge-invariant Higgs spectral function which is not plagued by positivity violations or unphysical thresholds. Furthermore, the mass extracted from the gauge-invariant bound state is only logarithmically sensitive to the scale of new physics at one-loop order in contrast to its elementary counterpart.  
\end{abstract}

\maketitle

\section{Introduction}
The perturbative treatment of the Higgs boson is highly successful in order to describe high-energy processes at the LHC. 
The definition of electroweak observables, e.g., the mass and decay width of the Higgs, as well as the computation of cross sections are commonly derived in terms of properties of the elementary fields of the standard model Lagrangian.
However, this description, based on the convenient picture of spontaneous electroweak symmetry breaking, is paradoxical at various levels \cite{Elitzur:1975im,Gribov:1977wm,Singer:1978dk,Fradkin:1978dv,Osterwalder:1977pc,Caudy:2007sf,Friederich:2011xs,Greensite:2018mhh}, for a review see \cite{Maas:2017wzi}.

In order to avoid field theoretical inconsistencies which we will briefly sketch in a moment, we will use a strict gauge-invariant definition of the Higgs particle/resonance that was first introduced by Fr\"ohlich, Morchio, and Strocchi (FMS) \cite{Frohlich:1980gj,Frohlich:1981yi}. They proposed to define electroweak observables as properties of gauge-invariant bound state operators  in a similar spirit to hadrons in QCD. At first sight, this seems to be at odds with any textbook knowledge about electroweak interactions. This seeming contradiction is resolved by the fact that the bound states of the non-Abelian electroweak theory precisely reproduce the usual results based on the elementary fields at leading order in the FMS framework. 
In particular, we will show that the pole structure of the gauge-invariant bound state propagator and the elementary Higgs propagator coincides to all orders in a loop expansion. Higher-order contributions of the FMS expansion result in loop suppressed modifications of off-shell quantities which further substantiates the usefulness of the common procedure. Nonetheless, enhancements above the two-Higgs threshold might be observable in future precision measurements. Furthermore, the higher-order FMS terms ensure a gauge-invariant, positive spectral function for the Higgs bound state.

Although different approaches exist in the literature to demonstrate gauge-parameter invariance of various quantities describing physical observables, these statements do not necessarily imply full gauge-invariance which is a stronger statement. 
Within the setting of perturbative gauge field theories, the BRST symmetry is used to construct the physical (perturbative) gauge-invariant state space. However, the BRST construction breaks down from a comprehensive perspective due to the Gribov-Singer ambiguity in a general non-Abelian gauge theory \cite{Gribov:1977wm,Singer:1978dk,vanBaal:1991zw,vanBaal:1997gu} albeit the precise manifestation and implications for a theory with Brout-Englert-Higgs (BEH) mechanism are not fully understood yet \cite{Lenz:2000zt,Maas:2010nc,Capri:2013oja}. 
Furthermore, Elitzur's theorem proves that the conventional picture of spontaneous gauge symmetry breaking is inadequate in general \cite{Elitzur:1975im}. From a technical perspective, the breaking is merely caused by the gauge-fixing procedure and not the form of the Higgs potential.

As a result, any nontrivial structure of the elementary Higgs propagator is solely induced by the gauge fixing in a continuum formulation of the theory. Strictly speaking, the gauge fixing introduces a nontrivial coupling between the boundary conditions and the expectation values of gauge dependent operators as can be shown in gauge-fixed lattice formulations \cite{Frohlich:1980gj,Frohlich:1981yi}. Indeed, the propagator of the elementary Higgs field has a trivial form within a lattice approach where no gauge fixing is performed \cite{Maas:2017wzi}. In general any Green's function of a gauge dependent object such as the Higgs or the gauge bosons will vanish as long as an action and measure invariant under local gauge transformations is used.

These considerations immediately lead to an apparent contradiction. On the one hand, we have the tremendous success of the perturbative continuum formulation to explain current and past collider experiments. On the other hand, basic field theoretical arguments call the validity of the conventional approach into question. The solution to this problem was found by the pioneering work of Fr\"ohlich, Morchio, and Strocchi via their aforementioned reformulation of electroweak observables as properties of gauge-invariant bound state operators \cite{Frohlich:1980gj,Frohlich:1981yi}. Additionally, the FMS approach provides the field theoretical foundation why the standard perturbative gauges and the investigations of gauge-dependent elementary fields are able to reflect the underlying physics appropriately.

That this is possible can be traced back to two points. First, the usual class of $R_{\xi}$ gauges allows for a nonvanishing vacuum expectation value (VEV) of the scalar doublet. Doing such a gauge fixing provides a mapping from a gauge-invariant bound state operator to an object of the remaining symmetry group. 
Second, from a group theoretical point of view, the weak sector of the standard model is special as it contains besides the non-Abelian $\SU(2)_{\mathrm{gauge}}$ gauge structure a global $\SU(2)_{\mathrm{global}}$ symmetry acting nonlinearly on the scalar doublet. Within the conventional picture, both groups are broken to a global diagonal subgroup $\SU(2)_{\mathrm{gauge}} \times \SU(2)_{\mathrm{global}} \to \SU(2)_{\mathrm{diag}}$. The fact that these groups coincide allows a one-to-one mapping of gauge-invariant bound state operators and elementary fields within the electroweak standard model at leading order in the FMS formalism.\footnote{As many beyond the standard model theories with an extended BEH sector do not fulfill this special feature, the duality relation imposed by the FMS mechanism is different and may cause a qualitative different spectrum \cite{Sondenheimer:2019idq,Maas:2017xzh,Maas:2016ngo,Maas:2017pcw,Maas:2018xxu}.}

Within the perturbative description, the original gauge structure is then partly encoded in the BRST symmetry which manifests in the Nielsen or Slavnov-Taylor identities. For instance, the Nielsen identities ensure that the pole mass of the elementary Higgs field is independent of the gauge-fixing parameter \cite{Nielsen:1975fs,Gambino:1999ai}. But this does not imply full gauge invariance \cite{Nielsen:1975fs}. It merely shows the gauge (parameter) independence of certain quantities within this particular class of gauges. As a counterexample one may consider the weak sector of the standard model in the temporal gauge. In this gauge, it can be shown that the VEV of the scalar field vanishes even if the scalar potential is of Mexican-hat type. The ``restoration of symmetry'', i.e., $\langle\phi\rangle = 0$, can be understood from the impact of topological defects. Common perturbative techniques are blind to these nonperturbative field configurations with nontrivial topology but their importance can be proven within nonperturbative approaches even at weak coupling \cite{Frohlich:1981yi}. Furthermore, not all quantities of interest are constrained by the virtue of the Nielsen or other identities. Although it can be shown that the pole of the propagator is gauge-parameter free, its residuum is not. Also other properties of the elementary Higgs propagator depend on the gauge. Further, the Lehmann-K\"all\`{e}n spectral density is in general not positive definite which prohibits a physical interpretation of the Higgs two-point function and demonstrates the limitations of the use of elementary fields. See Refs.~\cite{Dudal:2019aew,Dudal:2019pyg,Capri:2020ppe} for a recent analysis within an Abelian Higgs model.

Of course, all these points do not contradict the usefulness of the Nielsen identities within those gauges that provide a nonvanishing VEV, e.g., $R_{\xi}$ gauges, or the gauge-fixing procedure in general. 
In particular the results of the gauge-variant $n$-point functions of the elementary fields computed in common gauges will be of central importance in the FMS approach. Taking all such terms appearing on the right-hand side of the FMS expansion into account will allow us to enhance usual perturbative calculations to a gauge-independent framework operating on in principle nonperturbative bound states.

Although, these field theoretical facts are known for a long time, the actual phenomenological consequences of the higher-order terms induced by the FMS mechanism are barely investigated in the literature. Here, we will improve on this situation substantially.
In the following, we will analyze the structure of the bound state propagator of a gauge-invariant description of the Higgs. In Sec.~\ref{sec:Higgs}, we will first perform the FMS expansion to all orders. Then, we employ a consistent loop expansion to all terms obtained from the FMS expansion in Sec.~\ref{sec:one-loop}. After a brief recapitulation of the renormalization procedure and the (un)physical properties of the elementary Higgs propagator in Sec.~\ref{sec:elm-Higgs}, we show in Sec.~\ref{sec:Bound} that a gauge-invariant structure arranges in a natural way among the different terms of the FMS expansion albeit every individual term is gauge dependent. Subsequently, we demonstrate that phenomenological deviations from the usual treatment of the Higgs are present, but relatively small.

In Sec.~\ref{sec:lattice}, we benchmark the perturbative description of the gauge-invariant bound state operator via the FMS prescription with nonperturbative lattice simulations within the gauge-Higgs subsector of the standard model. Specifically, we test as to whether the perturbative treatment of the right-hand side of the FMS expansion is able to capture all relevant information of the bound state operator or under which circumstances nonperturbative effects might spoil the perturbative treatment.

\section{Gauge-invariant definition of the Higgs excitation and FMS mechanism}
\label{sec:Higgs}
A strict gauge-invariant formulation is desired for any observable of a gauge theory. Over the past years, overwhelming evidence has been accumulated that the scalar particle with a mass of ${\sim}$125 GeV observed in 2012 can be in some sense related to the standard model Higgs boson. It is directly affected by electroweak processes but only indirectly, e.g., via loop processes, to the strong interaction. Field theoretically, however, the excess in the invariant mass cross section cannot be caused by the elementary Higgs field as this field necessarily carries an unobservable non-Abelian gauge charge and some of its properties depend on the gauge. Nevertheless, considering the Higgs sector of the standard model, we can straightforwardly build a scalar bound state operator from the usual scalar doublet $\phi$ which is invariant with respect to the standard model gauge group, namely $\phid\phi$.

In general, a nonperturbative method is required to analyze the properties of this genuinely nonperturbative object. In contrast to QCD, however, the electroweak sector is a non-Abelian gauge theory with a BEH mechanism which leads to a convenient simplification of this bound state operator according to FMS \cite{Frohlich:1980gj,Frohlich:1981yi}. Suppose the potential of the scalar doublet has nontrivial minima. Then, we introduce a shifted doublet $\varphi$ to investigate excitations around one of these minima as usual,
\begin{align}
 \phi(x) = \frac{v}{\sqrt{2}}\phi_{0} + \varphi(x)
 \label{eq:split}
\end{align}
where $v$ is the modulus of the field configuration that minimizes the Higgs potential and $\phi_{0}$ is some unit vector in gauge space. Within the standard model, all minima belong to the same gauge orbit. Picking one particular representative, e.g., the common choice $\phi_{0}^{a}=\delta^{a2},$ corresponds to a gauge choice already at the classical level. Thus, also the identification of $\langle\phi\rangle = \frac{v}{\sqrt{2}}\phi_{0}$, i.e., the identification of the minimizing field configuration with the VEV of the scalar field, corresponds to a gauge choice and is only meaningful within this particular gauge. After the gauge is fixed, we can extract the elementary Higgs field and the would-be Goldstone modes which mix with the longitudinal parts of the gauge bosons from the fluctuations $\varphi$ around the minimum. In a covariant formulation, they can be expressed as $h = \sqrt{2}\re(\phid_{0}\varphi)$ and $\breve{\varphi} = \phi - \re(\phid_{0}\phi)\phi_{0}$, respectively. In the following, we will use the standard $R_{\xi}$ gauge-fixing condition in which the minimum selection process is already implemented by hand when it comes to an actual calculation.

Choosing a gauge condition that implements a nonvanishing VEV for the scalar field, leads to a convenient treatment of electroweak bound state operators. Within such a gauge, the scalar operator $|\phi|^{2}$ can be rewritten as 
\begin{align}
 \phid\phi = \frac{v^{2}}{2} + v h + \varphid\varphi
 \label{eq:boundstate}
\end{align}
via the split defined in Eq.~\eqref{eq:split}. Viewing the right-hand side of Eq.~\eqref{eq:boundstate} as an expansion in the fluctuations over the characteristic scale of electroweak physics and ignoring the unimportant constant term, we find that at nontrivial leading order in the expansion parameter $\varphi/v$ the properties of the gauge-invariant operator $|\phi|^{2}$ are described by the gauge-dependent elementary Higgs field $h$. In the following, we will show that this term provides indeed the dominant contribution from the phenomenological viewpoint. However, the next-to-leading order term $|\varphi|^{2}$ is important to render the full Green's function structure gauge invariant. 

All physical information of the bound state operator $|\phi|^{2}$ is encoded in its $n$-point functions. Studying the FMS expansion of the operator at the level of the propagator, we are able to rewrite the connected two-point function,
\begin{align}
 \langle (\phid\phi)(x) \; (\phid\phi)(y) \rangle
 &=  v^{2} \langle h(x) \; h(y) \rangle  +  2v \langle h(x) \; (\varphid\varphi)(y) \rangle \notag \\
 &\quad +  \langle (\varphid\varphi)(x) \; (\varphid\varphi)(y) \rangle,
 \label{eq:Propagator}
\end{align}
where we have used that the propagators depend only on the distance $|x-y|$. Thus, the FMS approach reduces the problem of calculating the propagator of an involved bound state operator to the computation of connected $n$-point functions of elementary fields. In particular, this allows for a perturbative access to bound state information by performing a loop expansion on the right-hand side. Further note that on the right-hand side every individual term is gauge dependent and can only be defined within the specifically chosen gauge. However, the sum of all the terms has to be gauge invariant by construction. This will be explicitly demonstrated at the one-loop level in the next sections. 

Based on the FMS expansion of the scalar operator $|\phi|^{2}$, it is not surprising that the bound state propagator can be approximated by the elementary Higgs propagator at leading order. The two higher-order terms in Eq.~\eqref{eq:Propagator}, the three-point function ${\sim}v$ and the four-point function ${\sim}v^{0}$, are neglected within conventional calculations. Their computation can be done with similar effort. In the following, we will concentrate on their importance regarding physical properties of the Higgs boson.  

Switching to momentum space, we parametrize the three terms on the right-hand side of Eq.~\eqref{eq:Propagator} as follows. The connected Higgs two-point function can be written as a geometric series where terms induced by quantum corrections are encoded in the self-energy function $\Sigma_{0}$, 
\begin{align}
 \langle h(p) h(-p)\rangle &= \frac{\I}{p^{2}-\mh^{2} - \Sigma_{0}(p^{2})}. 
\label{eq:FMSo0}
\end{align}
The subscript $0$ at the Higgs self energy $\Sigma_{0}$ denotes that this term originates from the leading order term with respect to the FMS expansion. In general, $\Sigma_{0}$ not only contains one-particle-irreducible (1PI) contributions but also tadpole diagrams attached to the propagators due to the breaking of local gauge invariance via gauge fixing. These emerge if the split~\eqref{eq:split} of the scalar field does not properly take the quantum corrections to the Higgs VEV into account, e.g., by choosing a minimum of the classical instead of the quantum effective potential as an expansion point. In such a case, the tadpole diagrams effectively shift the masses of the elementary fields accounting for these loop effects and contribute as momentum-independent terms that can be trivially included in the self energy. 
Performing the split at the minimum of the quantum effective potential removes these contributions. Of course, this can always be achieved by suitable renormalization conditions. In a scheme for which $\langle h \rangle = 0$; i.e., the VEV is fixed to its classical value, $\Sigma_{0}$ contains the standard 1PI diagrams only. Leaving the scheme for the moment unspecified, we will approximate $\Sigma_{0}$ by all diagrams up to a given loop order that are 1PI plus corresponding tadpole attachments, cf. Eq.~\eqref{eq:1loop0} for a one-loop approximation.

\begin{figure}
\centering
\begin{minipage}{0.1\textwidth}
 $ \langle h\, h \rangle$:
\end{minipage}
\begin{minipage}{0.1\textwidth}
\includegraphics[width=2.7cm]{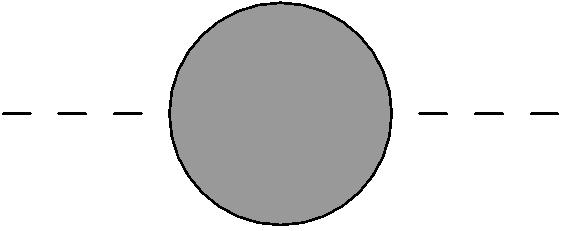}
\end{minipage}
\hfill \begin{minipage}{0.05\textwidth} \quad \end{minipage}\\[0.5cm]
\begin{minipage}{0.08\textwidth}
 $ \langle h\,\, |\varphi|^{2} \rangle$:
\end{minipage}
\begin{minipage}{0.2\textwidth}
\includegraphics[width=3.25cm]{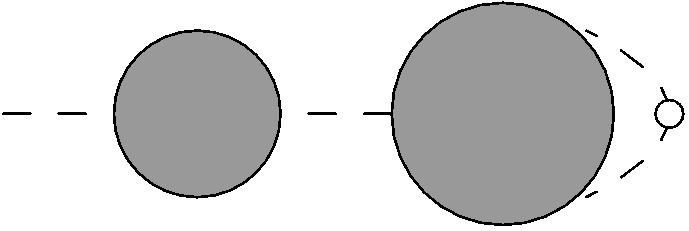}
\end{minipage}
+
\hfill
\begin{minipage}{0.1\textwidth}
\includegraphics[width=2.25cm]{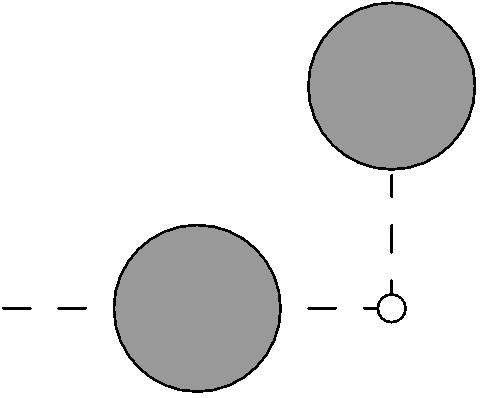}
\end{minipage}
\begin{minipage}{0.05\textwidth} \quad \end{minipage}\\[0.5cm]
\begin{minipage}{0.08\textwidth}
$\langle |\varphi|^{2} \,\, |\varphi|^{2} \rangle$: 
\end{minipage}
\begin{minipage}{0.125\textwidth}
\includegraphics[width=1.75cm]{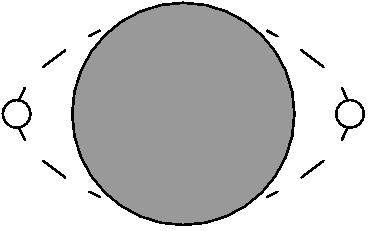}
\end{minipage}
\hspace{-0.2cm}+\,
\begin{minipage}{0.15\textwidth}
\includegraphics[width=4.5cm]{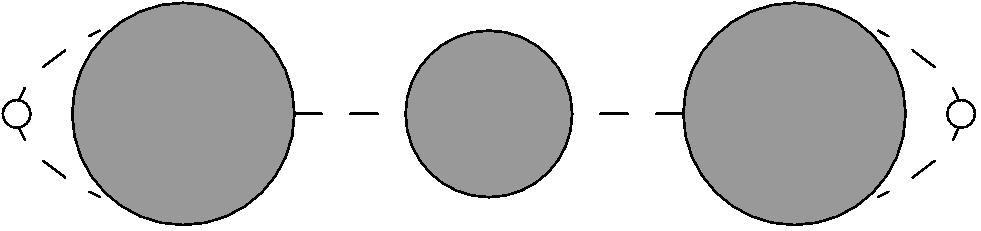}
\end{minipage}
\begin{minipage}{0.1\textwidth} \quad \end{minipage}\\[0.1cm]
\begin{minipage}{0.08\textwidth}
 \quad 
\end{minipage}
\begin{minipage}{0.03\textwidth} $+2$\end{minipage}
\begin{minipage}{0.175\textwidth}
 \includegraphics[width=3.25cm]{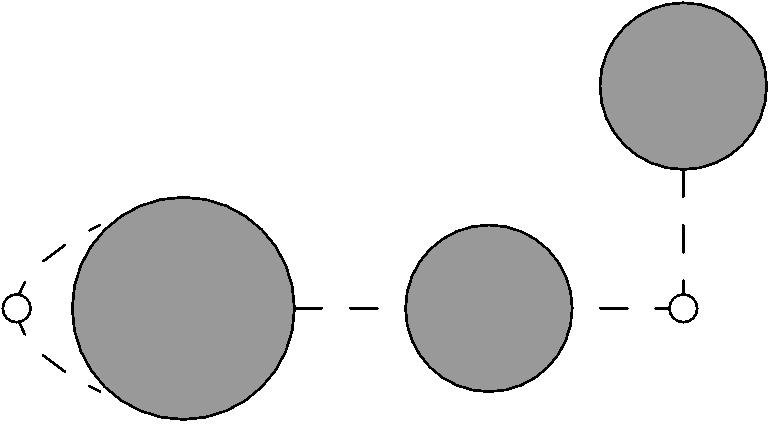}
\end{minipage}
\begin{minipage}{0.02\textwidth} $+$ \end{minipage}
\begin{minipage}{0.15\textwidth}
 \includegraphics[width=2.5cm]{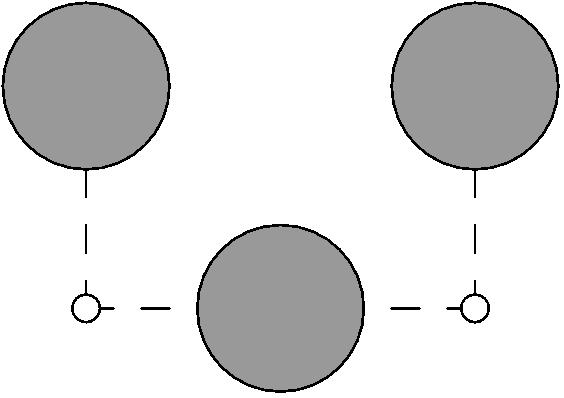}
\end{minipage}
\begin{minipage}{0.01\textwidth} \quad \end{minipage}
\caption{Diagrammatic sketch of the three Green's functions appearing due to the FMS expansion of the propagator of the gauge invariant operator $|\phi|^{2}$, see Eq.~\eqref{eq:Propagator}. Small white blobs denote $|\varphi|^{2}$ insertions. Gray circles encode tree level plus all possible loop corrections. For instance, a gray circle with one or two dashed lines attached indicates the sum of all tadpole diagrams or the full elementary Higgs propagator, respectively. Amputating the full external Higgs propagators at the two diagrams in the second line, i.e., the diagrammatic representation of $\langle h\,\, |\varphi|^{2} \rangle$, gives the two contributions to $\Sigma_{1}$ which are denoted by $\gamma^{(1,1)}$ and $\langle h \rangle$.}
\label{fig:diagrams-schematically}
\end{figure}

The three-point function appearing at next-to-leading order in the FMS expansion, can be viewed as the sum of all tadpole diagrams with suitable insertions of the composite operator $\varphid\varphi$. It will be convenient to define 
\begin{align}
 \langle h(p) \, (\varphid\varphi)(-p) \rangle  =  \frac{\I}{p^{2}-\mh^{2}-\Sigma_{0}(p^{2})} \Sigma_{1}(p^{2}),
\label{eq:FMSo1}
\end{align}
where the loop-induced function $\Sigma_{1}$, encoding the amputated contributions, can be further divided into two subgroups of diagrams, $\Sigma_{1} = \gamma^{(1,1)} + \langle h \rangle$, see Fig.~\ref{fig:diagrams-schematically}. Within the second subgroup ($\langle h \rangle$), the (amputated) external $h$-field propagator is directly attached to the composite operator insertion $\frac{1}{2}h^{2}$ (note: $\varphid\varphi = \frac{1}{2}h^{2} + \breve{\varphi}^{\dagger}\breve{\varphi}$). Then, a tadpole diagram has to be attached to the remaining line of the operator insertion, see the second diagram in the second line of Fig.~\ref{fig:diagrams-schematically}. Thus, we obtain a momentum-independent contribution to $\Sigma_{1}$ from these diagrams given by $\langle h \rangle(0)$. 
The other subgroup contains those diagrams where the tadpole bubble contains at least one internal scalar line at which we have a $|\varphi|^{2}$ insertion giving a nontrivial momentum contribution. Using a scheme that implements $\langle h \rangle = 0$, we only have to consider the subgroup of diagrams that are 1PI with an appropriate operator insertion, i.e., $\Gamma^{(1,1)}(p^{2})$, where $\Gamma$ denotes the effective action (generating functional of all 1PI diagrams) and we define the upperscripts according to  
\begin{align}
 \Gamma^{(n_{1},n_{2})} =  \bigg(\frac{\delta}{\delta K} - v h -\frac{1}{2}v^{2}\bigg)^{n_{2}}  \bigg(\frac{\delta}{\delta h}\bigg)^{n_{1}} \Gamma \bigg|_{0} .
\end{align}
Here, we have introduced $K$ as the source term for the gauge-invariant composite operator $|\phi|^{2}$; i.e., we modify the action according to $S_{\mathrm{SM}} \to S_{\mathrm{SM}} + \int_{x} K\, \phid\phi$. Thus, the operator $(\delta_{K} -vh -v^{2}/2)$ generates the required composite operator insertions $|\varphi|^{2}$. In a general scheme where the 1PI tadpole diagrams $\langle h \rangle_{1\mathrm{PI}} \equiv \Gamma^{(1,0)}$ are nonvanishing, we also have to take tadpole diagrams attached to the internal lines of the diagrams generated by $\Gamma^{(1,1)}$ into account. We label these contributions by $\gamma^{(1,1)}$. Thus, choosing the renormalization condition $\Gamma^{(1,0)} = 0$ will simplify the occurring structures as we have $\Sigma_{1} = \Gamma^{(1,1)}$ in this particular case. Using this notation, we may express $\gamma^{(2,0)} = p^{2}-\mh^{2}-\Sigma_{0}$ as the full inverse propagator including tadpole graphs. Correspondingly, $\Sigma_{0} = \Gamma^{(2,0)} - p^{2} + \mh^{2}$ if $\Gamma^{(1,0)} = 0$ is implemented.

In the same spirit, we may view the four-point function
\begin{align}
\!\!\! \langle (\varphid\varphi)(p) \, (\varphid\varphi)(-p) \rangle  \!&=  \I \Sigma_{2}(p^{2}) \notag\\
 &= \gamma^{(0,2)}(p^{2}) + \big(\Sigma_{1} \langle h\,h \rangle \Sigma_{1}\big)(p^{2})
 \label{eq:FMSo2}
\end{align}
as the sum of all vacuum bubbles with at least one internal scalar line to account for two insertions of the composite operator $|\varphi|^{2}$. The subscripts at $\Sigma_{1}$ and $\Sigma_{2}$ indicate that these loop contributions appear at next-to-leading and next-to-next-to-leading order of the FMS expansion, respectively. Also the function $\Sigma_{2}$ can be divided into two contributions originating from different classes of diagrams, see the second line of Eq.~\eqref{eq:FMSo2}. The first class described by $\gamma^{(0,2)}$ reduces to all diagrams generated by $\Gamma^{(0,2)}$ if the renormalization condition $\Gamma^{(1,0)}=0$ is chosen. If not, it encodes all diagrams that are 1PI with two $|\varphi|^{2}$ insertions plus potential tadpole diagrams attached to internal propagators as in the previous cases. The second class is a non-1PI contribution where we find the operator insertions separated in two distinct structures given by $\Sigma_{1}$ where both are connected via an internal propagator. A diagrammatic representation can be found in the third and fourth row of Fig.~\eqref{fig:diagrams-schematically}. There, the first diagram describes $\gamma^{(0,2)}$ while the remaining three diagrams are given by $(\Sigma_{1})^{2}\langle h\,h \rangle$. 
Further note that only the first diagram contains a one-loop contribution while the remaining diagrams start at two-loop order. 
Implementing $\Gamma^{(1,0)} = 0$ gives $\I\Sigma_{2}(p^{2}) = \Gamma^{(0,2)}(p^{2}) + \big(\Gamma^{(1,1)}(p^{2})\big)^{2}\big(\Gamma^{(2,0)}\big)^{-1}(p^{2})$.

\section{One-loop approximation}
\label{sec:one-loop}
\begin{figure*}
\centering
\includegraphics[width=0.92\textwidth]{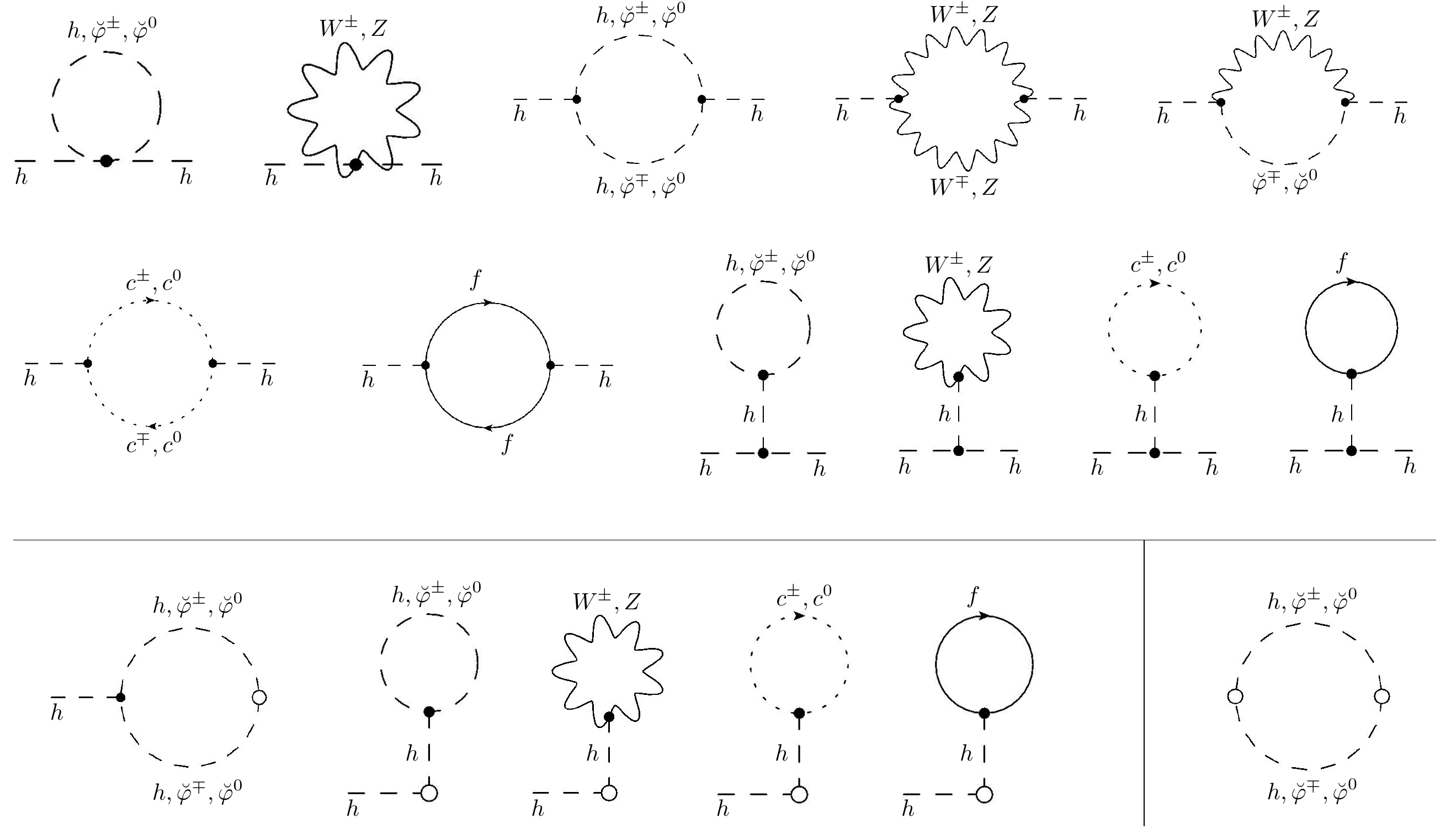}
\caption{Diagrammatic representation of $\Sigma_{0}$, $\Sigma_{1}$, and $\Sigma_{2}$ at one-loop order given in Eqs.~\eqref{eq:1loop0}-\eqref{eq:1loop2}. The one-loop contributions to the Higgs self-energy $\Sigma_{0}$ are given in the first two rows. The first five diagrams in the last row depict $\Sigma_{1}^{1\rml}$ while the last diagram shows $\Sigma_{2}^{1\rml}$.
Small white blobs denote $\varphid\varphi$ insertions.}
\label{fig:diagrams-1loop}
\end{figure*}

After these general formal considerations, we will restrict actual computations to the one-loop level throughout this paper. Using dimensional regularization, we obtain for the three unrenormalized functions $\Sigma_{0}$, $\Sigma_{1}$, and $\Sigma_{2}$ within this approximation:
\begin{widetext}
\begin{align}
 \Sigma_{0}^{1\rml}(p^{2}) 
 &= \frac{\varGamma\big(1-\frac{d}{2}\big)}{(4\pi)^{\frac{d}{2}}} \frac{1}{v^{2}} 
 \left[ -3\mh^{d} - 4(d-1)\mW^{d} - 2(d-1)\mZ^{d} + 3d_{\gamma}\sum_{\mathrm{f}} N_{\rmC\rmf}\mf^{d} - 2p^{2}\mW^{d-2} - p^{2}\mZ^{d-2} \right]   \notag \\
 &\quad + \frac{1}{v^{2}} \bigg[ -\frac{9}{2} \mh^{4}\, \loopB(\mh^{2}) - \Big(p^{4}-4\mW^{2}p^{2}+4(d-1)\mW^{4}\Big)\, \loopB(\mW^{2})   \notag \\ 
 &\qquad\qquad  - \frac{1}{2}\Big(p^{4}-4\mZ^{2}p^{2}+4(d-1)\mZ^{4}\Big)\, \loopB(\mZ^{2}) + (d-1)d_{\gamma} \sum_{\mathrm{f}} N_{\rmC\rmf}\mf^{2}\, \loopF(\mf^{2}) 
  \bigg]   \notag \\
 &\quad + \frac{1}{v^{2}} \left[ \frac{\varGamma\big(1-\frac{d}{2}\big)}{(4\pi)^{\frac{d}{2}}} (p^{2}-\mh^{2}) \Big[ 2(\xi \mW^{2})^{\frac{d}{2}-1} + (\xi \mZ^{2})^{\frac{d}{2}-1} \Big] + (p^{4}-\mh^{4}) \Big[ \loopB(\xi\mW^{2}) +\frac{1}{2}\loopB(\xi\mZ^{2}) \Big] \right],
\label{eq:1loop0} 
\\[0.5cm]
 \Sigma_{1}^{1\rml}(p^{2}) &= \frac{\varGamma\big(1-\frac{d}{2}\big)}{(4\pi)^{\frac{d}{2}}} \frac{1}{2v} \bigg[ -3\mh^{d-2} - 4(d-1) \frac{\mW^{d}}{\mh^{2}} 
   - 2(d-1) \frac{\mZ^{d}}{\mh^{2}} + 2d_{\gamma} \sum_{\rmf} N_{\rmC\rmf}\frac{\mf^{d}}{\mh^{2}} \bigg]  
   - \frac{1}{2v} 3\mh^{2} \loopB(\mh^{2}) \notag \\
 &\quad\ +  \frac{\varGamma\big(1-\frac{d}{2}\big)}{(4\pi)^{\frac{d}{2}}} \frac{1}{2v} \bigg[ - 2 (\xi\mW^{2})^{\frac{d}{2}-1} - (\xi\mZ^{2})^{\frac{d}{2}-1} \bigg] + \frac{1}{2v} \mh^{2}\bigg[ - 2\loopB(\xi\mW^{2}) - \loopB(\xi\mZ^{2}) \bigg], 
\label{eq:1loop1}   
\\[0.5cm]
\Sigma_{2}^{1\rml}(p^{2}) &= -\frac{1}{2} \left[ \loopB(\mh^{2}) + 2\loopB(\xi\mW^{2}) + \loopB(\xi\mZ^{2}) \right].
\label{eq:1loop2} 
\end{align}

\end{widetext}
Here, we have used the abbreviations
\begin{align*}
 \loopB(m^{2}) &= \frac{\varGamma\big(2-\frac{d}{2}\big)}{(4\pi)^{\frac{d}{2}}} \int_{0}^{1} dx \, \Big(m^{2}-(x-x^{2})p^{2}\Big)^{\frac{d}{2}-2}, \\
 \loopF(m^{2}) &= -\frac{\varGamma\big(1-\frac{d}{2}\big)}{(4\pi)^{\frac{d}{2}}} \int_{0}^{1} dx \, \Big(m^{2}-(x-x^{2})p^{2}\Big)^{\frac{d}{2}-1}
\end{align*}
to encode one-loop integrals containing two bosonic or two fermionic lines, respectively. The sums over the index $f$ run over all fermion flavors and $N_{\rmC\rmf} =3$ for quarks and $N_{\rmC\rmf} = 1$ for leptons. For convenience, we have chosen the same gauge-fixing parameter for the $W$ and $Z$ gauge field, $\xi\equiv \xi_{\mathrm{W}} = \xi_{\mathrm{Z}}$. Of course, all following cancellation mechanisms work for distinct gauge-fixing parameters as well. A diagrammatic representation of the one-loop approximations can be found in Fig.~\ref{fig:diagrams-1loop}. The result for $\Sigma_{0}^{1\rml}$ can, of course, be found in the literature, e.g., see \cite{Denner:1991kt} where it is expressed in terms of the Passarino-Veltman functions. 
As we have not specified any renormalization scheme yet, we explicitly included tadpole graphs. In order to extract only the 1PI information, 
we may use the one-loop relations $\Sigma_{0}^{1\rml} = \I \Gamma^{(2,0)1\rml} - \frac{3}{v}\I\Gamma^{(1,0)1\rml} -p^{2}+\mh^{2}$ and $\Sigma_{1}^{1\rml} = \Gamma^{(1,1)1\rml} - \frac{1}{\mh^{2}}\I\Gamma^{(1,0)1\rml}$ where the tadpole contributions read,
\begin{align}
 \I \Gamma^{(1,0)1\rml} &= \frac{\varGamma\big(1-\frac{d}{2}\big)}{(4\pi)^{\frac{d}{2}}} \frac{1}{v} 
 \bigg[
  \frac{3}{2}\mh^{d} + 2(d-1)\mW^{d}   \notag \\
 &\quad+ (d-1)\mZ^{d} - d_{\gamma}\sum_{\mathrm{f}}\mf^{d}   \notag \\ 
 &\quad+ \mh^{2}(\xi\mW^{2})^{\frac{d}{2}-1} + \frac{1}{2}\mh^{2}(\xi\mZ^{2})^{\frac{d}{2}-1} \bigg].
\label{eq:tadpole}
\end{align}
Terms that explicitly depend on the gauge-fixing parameter $\xi$ are separated in the last lines of Eqs.~\eqref{eq:1loop0}, \eqref{eq:1loop1}, and \eqref{eq:tadpole} as well as in the last two terms of Eq.~\eqref{eq:1loop2}. 
For later purpose, we also define quantities with a tilde which denote the original quantity minus any contribution coming from modes that carry a $\xi$-dependent mass term, i.e., $\tilde\Sigma_{0}^{1\rml} = \Sigma_{0}^{1\rml} -$[last line of Eq.~\eqref{eq:1loop0}], $\tilde\Sigma_{1}^{1\rml} = \Sigma_{1}^{1\rml} -$[last line of Eq.~\eqref{eq:1loop1}], $\tilde\Sigma_{2}^{1\rml} = -\frac{1}{2}\loopB(\mh^{2})$.

\section{Elementary Higgs propagator}
\label{sec:elm-Higgs}
Before we discuss the properties of the gauge-invariant bound state operator in detail, we briefly recapitulate some properties of the elementary Higgs propagator.

\subsection{Renormalization}
\label{sec:ren-el}
First, we have to introduce counterterms to absorb the UV divergencies in the $d\to 4$ limit. As dictated by the general perturbative renormalization procedure, the divergent parts only appear in powers of the external momentum $p^{2}$. At the level of the two-point function, we expect terms being ${\sim} p^{0}c_{d-4}$ and ${\sim} p^{2}c_{d-4}$, where $c_{d-4} = \frac{2}{d-4} - \gamma_{\mathrm{E}} + \ln(4\pi)$, which can be absorbed by the mass and wave function counterterms, respectively. Studying the one-loop contributions to $\Sigma_{0}$ this is certainly the case. At first sight, it seems that additional divergent terms ${\sim}p^{4}$ appear due to gauge boson loops, e.g., $p^{4}\loopB(\mW^{2})$, $p^{4}\loopB(\xi\mW^{2})$, $\cdots$. However, the sum of all these contributions is UV finite, i.e., the $c_{d-4} p^{4}$ dependence cancels precisely for any $\xi$. This is expected as we work in a perturbatively renormalizable gauge. As a side remark, the whole term ${\sim}p^{4}$ is absent in case of $\xi=1$ (Feynman--'t Hooft gauge). 
Summarizing, the divergent part of $\Sigma_{0}^{1\rml}$ reads
\begin{align*}
\Sigma_{0,\mathrm{div}}^{1\rml} = \frac{c_{d-4}}{16\pi^{2}v^{2}} \bigg\{& \Big[(3-\xi)(2\mW^{2}+\mZ^{2}) - \frac{d_{\gamma}}{2}\sum_{f}\mf^{2}\Big]p^{2} \\
 & - \Big[3\mh^{4} - (2\xi\mW^{2} + \xi\mZ^{2})\mh^{2} \Big] \bigg\}.
\end{align*}

The divergent term quadratic in the external momentum can be removed by the renormalization parameter of the scalar kinetic counterterm $\delta_{Z}|\partial_{\mu}\phi|^{2}$. For the momentum independent part, we have more freedom due to the breaking of the gauge structure via gauge fixing. In the pure scalar sector, this breaking technically manifests in the same way as for a scalar field theory with spontaneously broken global symmetry. Thus, the mass and quartic coupling counterterm $\delta_{m} \phid\phi - \frac{\delta_{\lambda}}{2} (\phid\phi)^{2}$ can be fixed via various Green's functions as can be seen by inserting the split~\eqref{eq:split} into the counterterm Lagrangian. Then, the counterterm Lagrangian of the pure Higgs sector reads,
\begin{align}
 \mathcal{L}_{\mathrm{H,c.t.}} 
 &= \left(\delta_{m}-\frac{v^{2}}{2}\delta_{\lambda}\right)v h \notag \\
 &\quad + \frac{1}{2}\left(\delta_{m}-\frac{3v^{2}}{2}\delta_{\lambda}\right)h^{2} 
  + \left(\delta_{m}-\frac{v^{2}}{2}\delta_{\lambda}\right) \breve{\varphi}^{\dagger}\breve{\varphi}
 \notag \\
 &\quad - \delta_{\lambda} v(h^{3} + 2h \breve{\varphi}^{\dagger}\breve{\varphi})
 - \frac{\delta_{\lambda}}{8} (h^{2}+2\breve{\varphi}^{\dagger}\breve{\varphi})^{2}
\label{eq:counterterm}
\end{align}
where we have omitted the kinetic term as it translates trivially. Independently which two of the $(n{\leq} 4)$-point functions of the Higgs excitation $h$ or the would-be Goldstone bosons $\breve{\varphi}$ we choose to determine the two renormalization parameters $\delta_{m}$ and $\delta_{\lambda}$, all of them will become UV finite due to the underlying symmetry. As we focus our analysis on the Higgs propagator as this is the leading order term of the FMS expansion and the only $n$-point function appearing on the right-hand side without an additional operator insertion, it will be most convenient for our purposes to fix one of the two parameters via $\langle h\,h\rangle$. Without loss of generality, we choose $\delta_{m}$ and leave $\delta_{\lambda}$ arbitrary for the moment. Using this choice and including the counterterms, the renormalized self-energy reads at the one-loop level,
\begin{align}
\Sigma_{0} &= \Sigma_{0}^{1\rml} - \delta_{Z}p^{2} -\left[ \delta_{m} - \frac{3}{2}v^{2}\delta_{\lambda}\right] +\frac{3}{v}\left(v\delta_{m}-\frac{1}{2}v^{3}\delta_{\lambda}\right) \notag\\
&= \Sigma_{0}^{1\rml} - \delta_{Z}p^{2} + 2\delta_{m}.
\end{align}
In the first line, we have explicitly illustrated the momentum-independent counterterm contributions coming from the 1PI diagram in square brackets, i.e., from the term ${\sim}h^{2}$ in Eq.~\eqref{eq:counterterm}, and from the one-particle reducible tadpole diagram [${\sim}h$ in Eq.~\eqref{eq:counterterm}] in parentheses. 

Of course, the removal of the UV divergent parts can be done via different renormalization strategies. Most common for high-order computations are the MS and $\overline{\mathrm{MS}}$ scheme such that the counterterms remove the pole ${\sim}\frac{1}{d-4}$ or the terms ${\sim}c_{d-4}$, respectively. For our purposes, however, a pole/on-shell scheme is more appropriate. Within these schemes, the finite parts of the counterterms are chosen in such a way that the renormalized parameters coincide with physical parameters. This can be achieved to all orders in perturbation theory. 

The on-shell scheme can be straightforwardly implemented for stable particles. However, the extraction of the mass and width of an unstable particle as the SM Higgs boson is involved in a gauge theory. In particular, it can be proven that the on-shell mass and width may become gauge-dependent in next-to-next-to-leading order of the perturbative expansion spoiling any physical interpretation of the two quantities \cite{Sirlin:1991rt,Stuart:1991xk,Stuart:1991cc,Aeppli:1993rs}. In order to circumvent this obvious problem, different strategies have been developed. A useful generalization is given by the complex mass scheme that can be viewed as an analytical continuation of the on-shell scheme by introducing complex renormalization constants and choosing the complex pole of the two-point function as a renormalization point \cite{Kniehl:1998fn,Denner:2005fg,Denner:2006ic,Bredenstein:2006rh,Kniehl:2008cj}. This provides a gauge-parameter independent renormalization condition for the mass of an unstable particle within the class of $R_{\xi}$ gauges as the complex pole does not depend on the gauge-fixing parameters. 

More precisely, we demand $\Sigma_{0}(p^{2}=\mh^{2}) = 0$ and $\Sigma_{0}'(p^{2}=\mh^{2}) = 0$, $\mh^{2} \in \mathbb{C}$, to implement the complex mass scheme where the prime denotes differentiation with respect to $p^{2}$.\footnote{Note that the imaginary parts of the renormalized parameters and the counterterms are related to render the bare parameters real valued and, thus, the action unitary. For instance, the mass counterterm has to fulfill the consistency condition $\im\, 2\delta_{m} = -\im\, \mh^{2}$. Further note that the Cutkosky cutting rules are not valid within this scheme \cite{Denner:2014zga}.} Defining $\mh^{2} = M_{\rmh}^{2} - \I M_{\rmh} \bold\Gamma_{\rmh}$, and considering the real and imaginary part of the renormalized self energy separately, the pole mass $M_{\rmh}$ and width $\bold\Gamma_{\rmh}$ can be obtained from
\begin{align}
 \re\, \Sigma_{0} (\mh^{2}) = 0 \quad \mathrm{and} \quad  M_{\rmh} \bold\Gamma_{\rmh} = - \im\,\Sigma_{0} (\mh^{2}).
 \label{eq:polescheme}
\end{align}
Using these two quantities to define the physical mass and width of the unstable Higgs particle provides a gauge-parameter independent definition within a perturbative approximation. 

An expansion of $\Sigma_{0}(\mh^{2})$ around $M_{\rmh}^{2}$ in Eq.~\eqref{eq:polescheme} and considering only terms up to $\mathcal{O}(\bold\Gamma_{\rmh})$ leads to the definition of the on-shell mass and width. 
Thus, the on-shell quantities might be viewed as the narrow width approximation of the pole quantities. 
In contrast to the complex mass scheme, the on-shell scheme is defined by 
\begin{align}
 \re\big[\Sigma_{0}(p^{2}=\mh^{2})\big] = 0, \quad \re\big[\Sigma'_{0}(p^{2}=\mh^{2})\big] = 0
\end{align}
for real-valued $\mh^{2} = M_{\rmh}^{2} \in \mathbb{R}$. Implementing this scheme, we obtain for the inverse resummed elementary Higgs propagator at the one-loop level
\begin{align}
\I \langle h\, h \rangle^{-1}(p^{2}) &=  \Big(1+ \re\,\Sigma_{0}^{1\rml \, \prime}(\mh^{2})\Big) (p^{2}-\mh^{2}) \notag \\
&\quad-  \Sigma_{0}^{1\rml}(p^{2}) + \re\,\Sigma_{0}^{1\rml}(\mh^{2}).
\label{eq:renProp}
\end{align}
The extraction of either the mass for a stable particle ($\Sigma_{0}(\mh^{2}) \in \mathbb{R}$) or of the mass and width of an unstable particle is straightforward within this scheme. The physical mass can be identified with the parameter $\mh=M_{\rmh}$ and the width is encoded in the imaginary part of the self energy,
\begin{align}
 \bold{\Gamma}_{\rmh} = - \frac{1}{\mh}\frac{\im\, \Sigma_{0}(\mh^{2})}{1-\re\, \Sigma_{0}'(\mh^{2})}.
 \label{eq:width}
\end{align}
As the ratio $\bold\Gamma_{\rmh}/M_{\rmh}$ is indeed small for the standard-model Higgs boson, Eqs.~\eqref{eq:renProp} and \eqref{eq:width} are sufficiently good approximations for its description at one-loop order.

\subsection{Gauge-(in)dependent properties}
As one can directly infer from the one-loop approximation, the pole of the Higgs propagator is gauge-parameter independent at lowest order in the perturbative expansion as $\partial_{\xi}\Sigma_{0}^{1\rml}(\mh^{2}) = 0$. That this statement holds to all orders in the perturbative expansion is ensured by the Nielsen identities \cite{Nielsen:1975fs,Gambino:1999ai}. This might lead to the conclusion that the mass extracted from the Higgs propagator is gauge-invariant and thus a physical observable. Nevertheless, we would like to emphasize at this point that the Nielsen identities merely show the gauge-parameter independence within a specific (and widely used) class of gauges. Other gauges than $R_{\xi}$ gauges can be chosen such that the elementary Higgs field propagator does not show such particlelike properties and nonperturbative methods are required to extract physical information of the system \cite{Frohlich:1981yi}. Furthermore, the residue of the pole of the Higgs is in general not gauge independent. This is in contrast to the massive elementary gauge bosons where it can be shown that the residues of the $Z$ and $W^{\pm}$ bosons are gauge parameter free to all orders and a gauge parameter invariant definition of partial widths is possible \cite{Grassi:2001bz}. For the Higgs this is not the case as can be directly inferred from the one-loop approximation, i.e., $\partial_{\xi} {\Sigma_{0}^{1\rml}}'(\mh^{2}) \neq 0$. A detailed discussion of this fact was done within the context of an Abelian-Higgs model \cite{Dudal:2019aew}. However, this is not a problem at a practical level as this gauge-parameter dependence does not contribute to any physical $S$-matrix element \cite{Haussling:1996rq}. Moreover, it is possible to fix a scheme such that the residue becomes $\xi$ independent by absorbing the $\xi$ dependence in the wave function counterterm as in the complex mass scheme.

\begin{figure}
\centering
\includegraphics[width=0.48\textwidth]{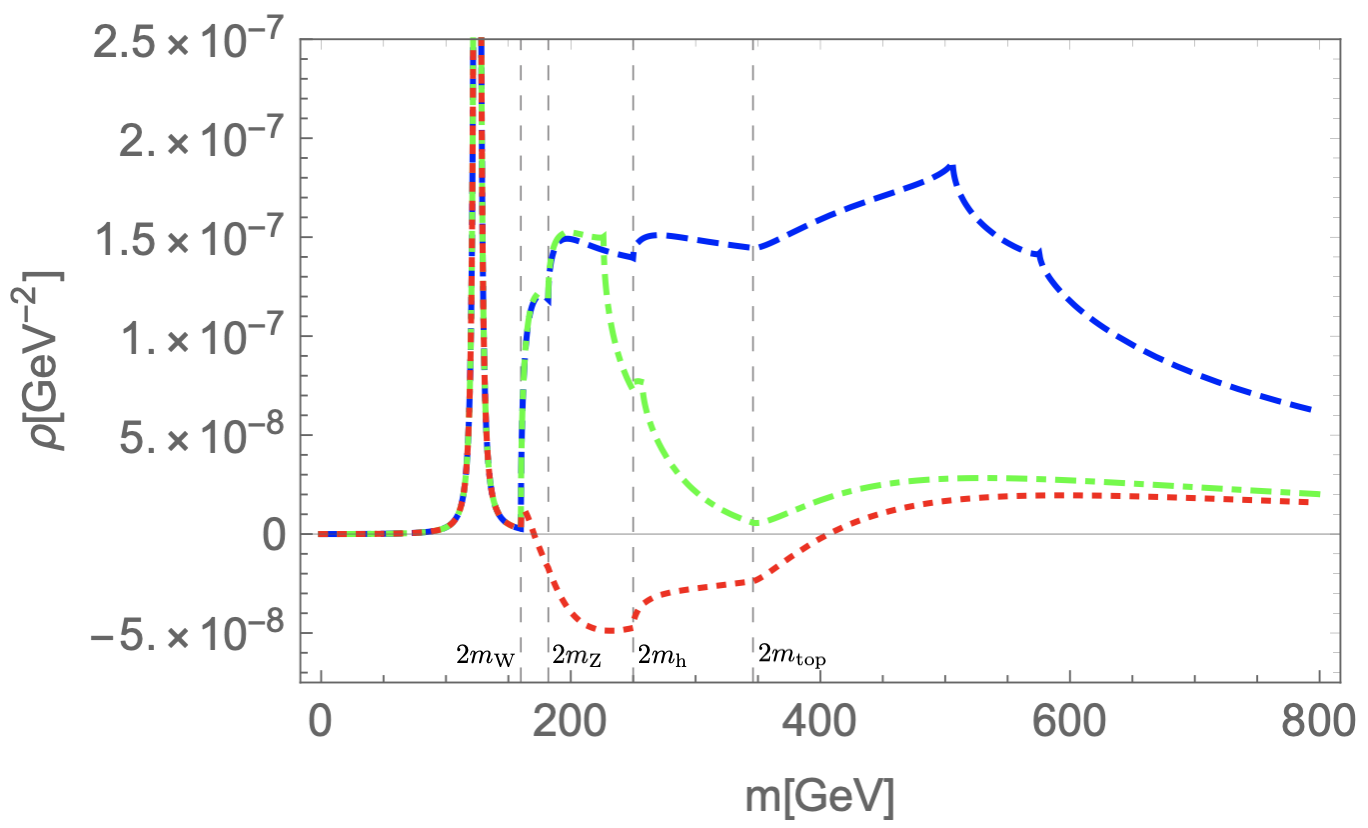}
\caption{Spectral density of the elementary Higgs field for different values of the gauge-fixing parameter $\xi$. We depict the spectral function for $\xi=1$ (red dotted line), $\xi=2$ (green dash-dotted line), and $\xi=10$ (blue dashed). The vertical gray dashed lines indicate the mass thresholds at $2\mW$, $2\mZ$, $2\mh$, and $2m_{\mathrm{top}}$ from left to right.}
\label{fig:spectral}
\end{figure}

In addition, other quantities that are extracted from the two-point function and would define the properties of a physical state are gauge dependent. For instance, let us briefly discuss the characteristics of the spectral density $\rho$ defined via
\begin{align*}
 \langle h(p)h(-p) \rangle = \int\limits_{0}^{\infty} \mathrm{d}m^{2} \frac{\rho(m^{2})}{p^{2}-m^{2}}
\end{align*}
which will obviously depend on the gauge-fixing parameter $\xi$. A detailed discussion of this feature can be found in Refs.~\cite{Dudal:2019pyg,Dudal:2019aew} for the Abelian Higgs model. Here, we observe similar properties albeit having a more involved structure due to the extended field content, see Fig.~\ref{fig:spectral}. 

First of all, we observe an expected peak at the Higgs mass whose width is dominated by the strength of the bottom Yukawa coupling.  
Further, we find nontrivial structures starting at the respective thresholds of the various observed particle masses.\footnote{Note that these physical scales need a gauge-invariant description as well. Again, the original FMS work showed that gauge-invariant bound states can be constructed for the electroweak gauge bosons as well as the standard model fermions which map to their elementary counterparts in the same spirit as for the Higgs. Thus, the mass scales set by the $W$ and $Z$ bosons as well as the leptons can be described in an appropriate fashion. For quarks the situations is more involved due to the additional non-Abelian color charge such that physical particle scales are set by meson and baryon masses. Taking such nontrivial hadronization effects into account is clearly beyond the scope of this work and we only remain with electroweak gauge-invariant structures in the following.} However, we find additional unphysical structures associated with fluctuations of mass $\sqrt{\xi} m_\mathrm{W/Z}$ which can be identified by varying the gauge fixing parameter $\xi$. This effect is most apparent for $\xi<1$. In this case, we obtain one-loop contributions already below the lowest physical mass threshold at $2\mW$ within the pure bosonic sector of the standard model. Even worse, for some values $\xi$ the spectral function becomes negative. Neglecting for a moment any fermionic fluctuations, we find that the spectral density $\rho$ becomes negative at large momentum for a sufficient small gauge-fixing parameter ($\xi<3$). This is in agreement with the situation in the Abelian Higgs model.\footnote{During the completion of this work, another analysis of the spectral function within an $\SU(2)$ Higgs model further substantiates this fact \cite{Dudal:2020uwb}.} Including fermionic contributions, top quark fluctuations alleviate nonpositivity issues in the large momentum regime. However, we still observe positivity violations at intermediate scales below $2m_{\mathrm{top}}$ for $\xi \lesssim 1.8$. 
For gauges with larger $\xi$, top contributions seem to cure positivity violations induced by longitudinal gauge boson and Goldstone modes. This is due to the fact that the dominant impact of these unphysical modes is only over a sufficient small momentum range in this case. Below $\sqrt{\xi}\mW$ they decouple while the strong top Yukawa coupling prevents the spectral density from becoming negative above the top threshold, e.g., compare the red dotted ($\xi=1$) and green dash-dotted line ($\xi=2$) in Fig.~\ref{fig:spectral}. 
Nevertheless, we have to state the following: although the elementary Higgs field propagator encodes some gauge parameter independent contributions, a full momentum dependent analysis reveals its unphysical nature spoiling any physical interpretation as a single particle due to the underlying non-Abelian gauge structure similar to the case of quarks and gluons within QCD.

Of course, the gauge-parameter dependence of the elementary Higgs propagator and the presence of unphysical thresholds is known for a long time. However, this seems not to be a problem at first sight. The Higgs is unstable within the standard model and occurs only as an intermediate resonance. Once a physical $S$-matrix element is calculated, the gauge-parameter dependence of internal Higgs propagators will be canceled by propagatorlike pieces of certain vertex and box diagrams \cite{Papavassiliou:1995fq,Papavassiliou:1995gs,Papavassiliou:1996zn}. Taking these processes into account, a $\xi$-independent definition of the Higgs self-energy can be defined via the pinch technique \cite{Papavassiliou:1997pb,Binosi:2009qm}. Neglecting any contributions with trivial external momentum dependence (tadpole and seagull diagrams), the Higgs self-energy computed via the pinch technique reads
\begin{align}
\Sigma_{0,\mathrm{pin}}^{1\rml} &= \frac{1}{v^{2}} \bigg[ -\frac{9}{2} \mh^{4}\, \loopB(\mh^{2})  \notag \\
 &\quad - \!\Big(\! \mh^{4} + 4\mW^{2} +3(d-1)\mW^{4} - 8\mW^{2}p^{2} \Big) \loopB(\mW^{2}) \notag\\
 &\quad -\frac{1}{2} \Big( \mh^{4} + 4\mZ^{2} +3(d-1)\mZ^{4} - 8\mZ^{2}p^{2} \Big) \loopB(\mZ^{2}) \notag \\
 &\quad + (d-1)d_{\gamma} \sum_{\mathrm{f}} N_{\rmC\rmf}\mf^{2}\, \loopF(\mf^{2})
 \bigg]. 
\end{align}
Obviously, the issue of unphysical thresholds of the propagator/spectral density will be cured by this strategy. In general, the pinch technique is a powerful tool to address perturbative gauge-invariance of calculations within non-Abelian gauge theories. Nonetheless, it still leaves some unsatisfactory open questions from a field theoretical perspective. First of all, the cancellation of the $\xi$-dependent terms via the involved interplay of propagator, vertex, and box diagrams works for any process where the elementary Higgs appears as an intermediate state. However, consider a model where the Higgs is a stable particle and the process where we have a Higgs as an asymptotic in and out state described by the (unpinched) elementary Higgs propagator. This object would still have the usual gauge-parameter dependence. Of course, this scenario is not realized within the standard model but the pinch technique Higgs propagator has another problem which is shared with its gauge-dependent counterpart. The spectral density still has negative contributions which appear in a similar fashion as for the $\xi=1$ case in Fig.~\ref{fig:spectral}. In the following section we will demonstrate that the gauge-invariant FMS formulation of the Higgs particle in terms of a bound state will solve this issue. Additionally, physical cross sections computed by the FMS formalism and the (pinch technique) elementary Higgs propagator are in good agreement for low momenta, which is consistent with the success of the latter method to describe present collider experiments.

\section{Bound state propagator}
\label{sec:Bound}
After this recapitulation of the elementary Higgs propagator, we will now consider the properties of the scalar bound state operator $\phid\phi$. First, we observe that the bound state propagator is gauge parameter independent up to one-loop order if we include all terms of the FMS expansion, see Eq.~\eqref{eq:Propagator}. More precisely, we restrict the $n$-point functions defined in Eqs.~\eqref{eq:FMSo0}, \eqref{eq:FMSo1}, and \eqref{eq:FMSo2} to their tree and one-loop representations and plug them into Eq.~\eqref{eq:Propagator}. Using the approximations~\eqref{eq:1loop0}-\eqref{eq:1loop2} for the three functions $\Sigma_{0}$, $\Sigma_{1}$, and $\Sigma_{2}$ and inserting unities of the form $(p^{2}-\mh^{2})^{n}/(p^{2}-\mh^{2})^{n}$ in front of the terms ${\sim}\Sigma_{n}$, we obtain
\begin{widetext}
\begin{align}
\langle (\phid\phi)(p) \, (\phid\phi)(-p) \rangle &= \frac{\I}{p^{2}-\mh^{2}} \notag \\ 
&\quad + \Bigg\{
 \frac{\varGamma\big(1-\frac{d}{2}\big)}{(4\pi)^{\frac{d}{2}}} 
 \Bigg[ -3 p^{2} \mh^{d-2}  - 2 p^{2} \bigg(1+2(d-1)\frac{\mW^{2}}{\mh^{2}}\bigg) \mW^{d-2} -  p^{2} \bigg(1+2(d-1)\frac{\mZ^{2}}{\mh^{2}}\bigg) \mZ^{d-2}  \notag \\
 &\qquad\qquad\qquad\qquad 
 + d_{\gamma} \sum_{\rmf} N_{\rmC\rmf} \bigg(\mf^{d} + 2p^{2} \frac{\mf^{d}}{\mh^{2}} \bigg) \Bigg]
 - \frac{1}{2}(p^{2}+2\mh^{2})^{2} \loopB(\mh^{2})  \notag \\
&\qquad\quad  - \Big(p^{4}-4\mW^{2}p^{2}+4(d-1)\mW^{4}\Big)\, \loopB(\mW^{2})  - \frac{1}{2}\Big(p^{4}-4\mZ^{2}p^{2}+4(d-1)\mZ^{4}\Big)\, \loopB(\mZ^{2})  \notag\\
&\qquad\quad + (d-1)d_{\gamma} \sum_{\mathrm{f}} N_{\rmC\rmf}\mf^{2}\, \loopF(\mf^{2})
  \Bigg\} \frac{\I}{(p^{2}-\mh^{2})^{2}}  + \mathcal{O}(\hbar^{2})  
\label{eq:Prop-Bound}
\end{align}
\end{widetext}
where we have separated the tree-level contribution in the first line while the one-loop contributions are given in the remaining lines. This important but trivial result was expected due to the construction of the bound state operator as a gauge charge singlet with respect to the standard model gauge group. Thus, all gauge-parameter dependent contributions of each single term on the right-hand side of the FMS expansion have to cancel within consistent (non)perturbative approximation schemes. In particular, we have $0 = \partial_{\xi} \langle |\phi|^{2}\, |\phi|^{2} \rangle = \partial_{\xi}( v^{2}\langle h\, h \rangle) + \partial_{\xi} (2v \langle h\, |\varphi|^{2} \rangle) + \partial_{\xi} \langle |\varphi|^{2}\, |\varphi|^{2} \rangle$. Going to some fixed order on the right-hand side, the gauge-parameter dependent terms have to be constrained by this identity, ensuring gauge-invariance once all terms of the FMS expansion are considered.

\subsection{Renormalization}
\label{sec:ren-bound}
Considering the three terms on the right-hand side of the FMS expansion, each term, the elementary Higgs propagator as well as the two terms with composite operator insertions, is separately UV divergent. Basically, we follow the standard formalism to renormalize the higher-order terms of the FMS expansion, i.e., $n$-point functions containing composite operators. However, a few comments are in order due to the symmetry breaking via gauge fixing that relates counterterms as in the case of the elementary field. 

A straightforward strategy to render the bound state propagator $\langle (\phid\phi) \; (\phid\phi) \rangle$ finite within the FMS approach is given by renormalizing each term appearing on the right-hand side of the FMS expansion individually. The renormalization procedure for the elementary Higgs propagator $\langle h \; h \rangle$ was discussed in Sec.~\ref{sec:ren-el}. For the next term in the expansion, $\langle h \; |\varphi|^{2} \rangle$, we may use an additional counterterm/wave function renormalization introduced in the source term accounting for a composite operator insertion $K \phid\phi \to Z_{|\phi|^{2}} K \phid\phi = (1+\delta_{|\phi|^{2}}) K \phid\phi$. Within a system with vanishing $v$, one would fix this additional counterterm by imposing a renormalization condition on $\langle \phi(x)\, \phid(y)\, |\phi|^{2}(z)\rangle$. In case the scalar field acquires a nonvanishing VEV, i.e., the $\mathrm{O}(4)$ symmetry of the pure Higgs sector is broken via gauge-fixing, we may use any \mbox{$n$-point} function containing the composite operator $\varphid\varphi$ and either two elementary Goldstone fields or up to two elementary Higgs fields, e.g., $\langle h(x) h(y) |\varphi|^{2}(z)\rangle$, $\langle \breve{\varphi}^{\pm/0}(x) \breve{\varphi}^{\mp/0}(y) |\varphi|^{2}(z)\rangle$, or $\langle h(x) |\varphi|^{2}(y)\rangle$. This is analogous to the determination of $\delta_{m}$ and $\delta_{\lambda}$ for the elementary Higgs propagator in the broken regime. Independently which option we choose, all other $n$-point functions with one composite operator insertion $|\varphi|^{2}$ are fixed and UV finite due to the underlying symmetry.

For our purposes it is obviously most convenient to choose $\langle h\; |\varphi|^{2}\rangle$. Actually, the counterterm $\delta_{|\phi|^{2}}$ only has to absorb the UV divergency of the 1PI part $\Gamma^{(1,1)}$ while the common tadpole counterterms take care of possible tadpole contributions attached to the 1PI diagrams if a scheme is used where $\langle h \rangle \neq 0$. Nevertheless, in order to fix the finite part of $\delta_{|\phi|^{2}}$, it does not matter if we impose the renormalization condition on $\Gamma^{(1,1)}$ or the function $\Sigma_{1}$ on a practical level as the finite parts of the tadpoles give only momentum independent contributions to any order in the loop expansion. A convenient choice is given by either $\re\, \Sigma_{1}(p^{2}=\mh^{2}) = 0$ or $\re\, \Gamma^{(1,1)}(p^{2}=\mh^{2}) = 0$ (on-shell scheme, $\mh\in\mathbb{R}$). Extending the complex mass scheme ($\mh\in\mathbb{C}$) to the case of composite operators, we could also choose $\Sigma_{1}(p^{2}=\mh^{2}) = 0$ or ${\Gamma^{(1,1)}(p^{2}=\mh^{2}) = 0}$. Of course, other schemes such as $\overline{\mathrm{MS}}$ are feasible as well.

Performing a one-loop approximation to make these considerations more explicit, i.e., $\langle h \; |\varphi|^{2} \rangle  =  \frac{\I}{p^{2}-\mh^{2}} \Sigma_{1}^{1\rml} + \mathcal{O}(\hbar^{2})$, we obtain,
\begin{align}
 \Sigma_{1} &= \Sigma_{1}^{1\rml} + v \delta_{|\phi|^{2}} + \frac{v}{\mh^{2}}\Big(\delta_{m}-\frac{v^{2}}{2}\delta_{\lambda}\Big) \notag\\
 &= \Gamma^{(1,1)1\rml} + v \delta_{|\phi|^{2}} - \frac{\I}{\mh^{2}} \bigg[ \Gamma^{(1,0)1\rml} + \I v\delta_{m} - \I\frac{v^{3}}{2}\delta_{\lambda} \bigg]. 
\end{align}
Here, we clearly see that the term in square brackets, coming from the tadpole contributions attached to the composite operator insertion, is already UV finite due to the renormalization of the $n$-point functions without any $|\varphi|^{2}$ insertion.
In the following, we will impose the renormalization condition $\re\, \Gamma^{(1,1)}(p^{2}=\mh^{2}) = 0$ within the on-shell scheme implying ${v\delta_{|\phi|^{2}} = - \re\, \Gamma^{(1,1)1\rml}(p^{2}=\mh^{2})}$. This is justified by the fact that we still have $\bold{\Gamma}_{h}/M_{h} \ll 1$ and we do not expect any gauge-dependent artifacts which usually spoil the applicability of the on-shell scheme at higher loop orders.

Finally, we have to discuss the renormalization of the highest order term of the FMS expansion, $\langle |\varphi|^{2} \, |\varphi|^{2} \rangle$. The function $\Sigma_{2}$ or rather $\Gamma^{(0,2)}$ is divergent already in free field theory. Therefore, we need an additional renormalization. Following our previous strategy, we impose the renormalization condition $\re\, \Gamma^{(0,2)}(p^{2} = \mh^{2}) = 0$ which is consistent with the tree level approximation. Thus, we have at the one-loop level ($\Sigma_{2}^{1\rml} = \Gamma^{(0,2)1\rml}$),
\begin{align}
 \Sigma_{2}(p^{2}) = \Sigma_{2}^{1\rml}(p^{2}) - \re\, \Sigma_{2}^{1\rml}(\mh^{2}).
\end{align}

This procedure is sufficient to renormalize all the structures by treating the gauge-invariant bound state propagator within the FMS approach. Therefore, we could proceed to extract physical information. Before doing so, however, we want to discuss a few particularities which are not present in other systems (due to the lack of the FMS expansion/BEH mechanism) by slightly changing the perspective on the calculation.

For this, let us plainly write down all counterterms say at the one-loop level and ignoring that they originated from three individual contributions.
\begin{widetext}
\begin{align}
\langle (\phid\phi)(p) \, (\phid\phi)(-p) \rangle &= \frac{\I v^{2}}{p^{2}-\mh^{2}} \notag + \frac{\I v^{2}}{p^{2}-\mh^{2}}\frac{\Sigma_{0}^{1\rml} - \delta_{Z}p^{2} + 2\delta_{m}}{p^{2}-\mh^{2}} \notag \\
&\quad + \frac{\I 2v}{p^{2}-\mh^{2}}\bigg(\Sigma_{1}^{1\rml} + v \delta_{|\phi|^{2}} + \frac{v}{\mh^{2}}\Big(\delta_{m}-\frac{v^{2}}{2}\delta_{\lambda}\Big) \bigg)
+\I \Big(\Sigma_{2}^{1\rml} - \tilde\delta_{|\phi|^{2}}\Big)
+ \mathcal{O}(\hbar^{2}) \notag \\
&= (\text{rhs. of Eq~\eqref{eq:Prop-Bound}}) + \frac{\I}{(p^{2}-\mh^{2})^{2}} \Bigg[
  (v^{4}\delta_{\lambda} - 2v^{2}\mh^{2} \delta_{|\phi|^{2}} - \mh^{4}\tilde\delta_{|\phi|^{2}})p^{0} \notag \\
 &\quad + \Big(-v^{2}\delta_{Z} + 2\frac{v^{2}}{\mh^{2}}\delta_{m} - \frac{v^{4}}{\mh^{2}}\delta_{\lambda} + 2v^{2}\delta_{|\phi|^{2}} + 2\mh^{2}\tilde\delta_{|\phi|^{2}} \Big)p^{2}  -  \tilde\delta_{|\phi|^{2}} p^{4} \Bigg], 
 \label{eq:Prop-Bound-Ren}
\end{align}
\end{widetext}
where we have introduced $\tilde\delta_{|\phi|^{2}}$ which accounts for the renormalization of $\langle |\varphi|^{2} \, |\varphi|^{2} \rangle$ within our previous consideration, i.e., we had $\tilde\delta_{|\phi|^{2}} = \re\, \Sigma_{2}^{1\rml}(\mh^{2})$ but leave this renormalization constant unfixed for the moment. 
From the unrenormalized result given in Eq.~\eqref{eq:Prop-Bound}, we know that we obtain UV divergencies of the form $c_{d-4}(d_{2}p^{4} + d_{1}p^{2} + d_{0})/(p^{2}-\mh^{2})^{2}$ at the one-loop order for the perturbative treatment of the FMS expanded bound state propagator. The renormalization constants originating from the split of the bare quantities into renormalized quantities and counterterms can be regrouped into the same momentum structure, see Eq.~\eqref{eq:Prop-Bound-Ren}. 
Therefore, we could equally impose renormalization conditions defining our scheme not in terms of the gauge-dependent Green's functions $\langle h\, h \rangle$, $\langle h\, |\varphi|^{2} \rangle$, and $\langle |\varphi|^{2}\, |\varphi|^{2} \rangle$ on the right-hand side of the expansion but on $n$-point functions of gauge-invariant physical (composite) fields, e.g., here the bound state propagator $\langle |\phi|^{2}\, |\phi|^{2} \rangle$. The latter carry the relevant information of the system as they are needed to define physical processes in a field-theoretical suitable manner. Thus from a pragmatical point of view, it is sufficient to properly renormalize the gauge-invariant $n$-point functions of the bound states as the elementary fields are anyhow only auxiliary quantities within our description.\footnote{Of course, this does not deny the usefulness of the elementary Higgs, weak gauge bosons, and left-handed fermions which play a similar role as quarks and gluons within QCD. None of them are physical observables but appropriate bound states with these auxiliary fields as constituents are. However, there is a difference to the strong interaction. Within QCD, it is most convenient to imposes renormalization conditions at the level of the elementary fields in order to perform any continuum calculation in a gauge-fixed setup and the actual bound state formation is an involved issue. In case of the weak sector, the BEH mechanism provides a duality relation explained by the FMS framework which links the elementary fields with the bound states in a completely different way such that the renormalization constants introduced for the gauge-dependent fields can straightforwardly be used to fix the gauge-invariant bound states within suitable gauges.} 
Taking this viewpoint, we would like to emphasize some particularities that appear at least at one-loop order but also discuss the question how this might extend beyond the one-loop approximation.

First, imposing renormalization conditions on the bound state propagator, e.g., via some modified on-shell/complex mass scheme, we obtain definitions of the finite parts of the renormalization constants that are gauge-parameter independent as any fluctuation induced contribution is gauge-parameter independent by definition, see Eq.~\eqref{eq:Prop-Bound}. This is in contrast to the above discussed strategy where we renormalize each (gauge-dependent) term appearing at different orders of the FMS expansion separately. Equivalently, we could do this at the level of the gauge-dependent elementary $n$-point functions by ignoring finite contributions that depend on $\xi$ throughout the renormalization process. As to whether this can be also done for the UV divergencies arising from these contributions is an interesting but intricate question. If so, we could completely avoid the necessity of considering Goldstone, ghost, and unphysical gauge boson modes and all their corresponding involved issues in order to extract physical information of a BEH model, although we still deal with an underlying non-Abelian gauge structure.

In fact, this strategy can be implemented at the level of the one-loop approximation but we have to be more careful at higher orders. 
In case all diagrams contributing to elementary $n$-point functions would only carry divergencies that are expected by power counting, i.e., UV divergencies ${\sim}p^{0}$ for all three functions $\Sigma_{0}$, $\Sigma_{1}$, and $\Sigma_{2}$ as well as a UV divergence ${\sim}p^{2}$ for $\Sigma_{0}$, the elementary counterterms would take care of any UV divergence and we could directly implement this scheme to any loop order. However, we also observed that some diagrams of the elementary Higgs propagator have a local divergence ${\sim}p^{4}$. This can be traced back to the presence of massive elementary vector bosons within the gauge-fixed formulation. Nonetheless, if we renormalize the $n$-point functions on the rhs. of the FMS expansion separately, the unusual divergencies originating from terms ${\sim}p^{4}\loopB(m_{\mathrm{W/Z}}^{2})$ were canceled by terms ${\sim}p^{4}\loopB(\xi m_{\mathrm{W/Z}}^{2})$ at the level of the one-loop elementary Higgs propagator as we work in a renormalizable gauge. Although the latter terms get canceled by higher-orders of the FMS expansion to ensure gauge-invariance such that the UV divergence cancellation ${\sim}p^{4}$ of $\Sigma_{0}$ is no longer intact, the fixing of $\tilde\delta_{|\phi|^{2}}$ by putting a renormalization condition on $\Sigma_{2}$ that contains $\loopB(\xi m_{\mathrm{W/Z}}^{2})$, and a reshuffling of the terms at the level of the bound state propagator guarantees UV finiteness.

Also in case we either fully ignore any contribution of loops that include internal lines of modes with $\xi$-dependent mass terms such that we do not obtain a cancelation between the divergencies of $p^{4}\loopB(m_{\mathrm{W/Z}}^{2})$ and $p^{4}\loopB(\xi m_{\mathrm{W/Z}}^{2})$ or we directly work at the level of the bound state propagator where the latter terms are not present, $\tilde\delta_{|\phi|^{2}}$ can be used to take care of the divergencies coming from the terms ${\sim}p^{4}\loopB(m_{\mathrm{W/Z}}^{2})$ at the one-loop level as can be seen in Eq.~\eqref{eq:Prop-Bound-Ren}. However, this will not be the case at higher orders where we obtain terms of the form $(p^{4}c_{d-4})^{n}/(p^{2}-\mh^{2})^{n+1}$ while $\tilde\delta_{|\phi|^{2}}=\tilde\delta_{|\phi|^{2}}(p^{2}-\mh^{2})^{n+1}/(p^{2}-\mh^{2})^{n+1}$. Such divergencies originate from terms of the form $\big(\tilde\Sigma_{0}^{1\rml}/(p^{2}-\mh^{2})\big)^{n}$ where $\tilde\Sigma_{0}^{1\rml}$ was defined at the end of Sec.~\ref{sec:one-loop} . Thus, the highest power of the external momentum in the numerator for the UV divergence and the renormalization constant coincides only in the $n=1$ loop case. 

Nevertheless, one should keep in mind that we considered so far only a subclass of non-1PI $n\geq 2$ loop diagrams and a plenty of further higher loop diagrams will contribute as well. 
It is an interesting question to examine if such divergencies that are canceled between $\xi$-dependent and $\xi$-independent loops for the two-point function $\langle h\, h \rangle$ such as $p^{4}c_{d-4}$ can be canceled by the higher-order terms of the FMS expansion in a strict gauge-parameter invariant setup and thus making any modes with $\xi$-dependent mass term obsolete or if these modes are needed to obtain a renormalizable field theory. At the one-loop level this can obviously be done but a general proof is beyond the scope of this simple analysis. Even if this is not the case, the FMS approach allows at least for a $\xi$-independent definition of the finite parts of the renormalization constants.\footnote{This extends also to a possible renormalization of the tadpole contributions. In order to simplify the computational effort, it will be convenient to force $\tilde{\Gamma}^{(1,0)}=0$ as we might anyhow ignore the $\xi$-dependent contributions to $\Gamma^{(1,0)}$ by the above strategy. Alternatively, one may fix $\delta_{\lambda}$ or $\delta_{m}$ to obtain $\Gamma^{(1,0)}=0$. However, this automatically implies that one of these renormalization constants becomes $\xi$ dependent which has to be reabsorbed by another renormalization parameter, e.g., $\delta_{|\phi|^{2}}$ or $\delta_{Z}$.}

Second, as the interplay of all terms of the FMS expansion is responsible for gauge-invariance, the counterterms originating from the different orders of the expansion become related. Thus, we need in fact only three of them instead of five to render the perturbative one-loop structures encoded in $\langle |\phi|^{2} |\phi|^{2} \rangle$ finite.

Third, the renormalization of the bound-state mass term shows an interesting behavior that is different from the renormalization of the mass parameter of the elementary field, at least at the one-loop level. Analyzing the counterterm structure ${\sim}p^{0}$ in Eq.~\eqref{eq:Prop-Bound-Ren}, we observe that it does not depend on $\delta_{m}$. Using a momentum cutoff regularization instead of dimensional regularization, $\delta_{m}$ is responsible for absorbing the quadratic cutoff dependence within the Higgs sector. 
More precisely, we find the usual $\Lambda^{2}$ dependency for the elementary Higgs self-energy, i.e.,
\begin{align}
 \Sigma_{0}^{1\rml} &= -\frac{3\Lambda^{2}}{8\pi^{2}v^{2}} \left(\mh^{2} + \mZ^{2} + 2\mW^{2} - \sum_{\rmf}\frac{N_{\rmC\rmf}d_{\gamma}}{3}\mf^{2}\right) \notag\\
 &\quad + \cdots
 \label{eq:cutoffReg0}
\end{align}
where $\cdots$ denote terms that are ${\sim}\ln\Lambda$ or ${\sim}\Lambda^{0}$. 
Including the higher-order terms of the FMS expansion for the bound state, we find that only $\delta_{\lambda}$, $\delta_{|\phi|^{2}}$, and $\tilde\delta_{|\phi|^{2}}$ which are $\sim \ln \Lambda$ contribute to the mass term renormalization, indicating that the renormalization of the bound state mass is only logarithmically sensitive to the cutoff scale. Indeed, we can explicitly show that terms quadratic in the UV cutoff get canceled by the higher-order terms of the FMS expansion being consistent with the counterterm structure. Going to the deep Euclidean region by neglecting mass thresholds and the impact of external momentum, we obtain at next-to-leading order of the FMS expansion
\begin{align}
 \Sigma_{1}^{1\rml} &= -\frac{3\Lambda^{2}}{16\pi^{2}v\mh^{2}} \left(\mh^{2} + \mZ^{2} + 2\mW^{2} - \sum_{\rmf}\frac{N_{\rmC\rmf}d_{\gamma}}{3}\mf^{2}\right) \notag \\ &\quad + \cdots.
 \label{eq:cutoffReg1}
\end{align}
Plugging Eq.~\eqref{eq:cutoffReg0} and Eq.~\eqref{eq:cutoffReg1} into the one-loop approximation for the bound state propagator and noting that $\Sigma_{2}^{1\rml}$ is only logarithmically divergent, we find that all $\Lambda^{2}$ dependencies precisely cancel, 
\begin{align}
v^{2} \Sigma_{0}^{1\rml} + 2v (-\mh^{2})\Sigma_{1}^{1\rml} + \mh^{4}\Sigma_{2}^{1\rml} = \mathcal{O}(\ln\Lambda).
\end{align}
Thus, the observable Higgs mass where we view the Higgs actually as the gauge-invariant bound state $\phid\phi$ is not quadratically sensitive to the scale of new physics at one-loop order from an effective field theory viewpoint. This is obviously different to the usual treatment where all properties of the Higgs are identified with the gauge-dependent elementary field whose infrared mass is extremely sensitive the bare mass parameter. However, accepting the point that the elementary field is rather a useful mathematical concept for particular classes of gauges and not an observable quantity and thus viewing experimental data from the gauge-invariant bound state viewpoint, it is not surprising that the mass runs logarithmically. The FMS approach allows us to treat $\langle |\phi|^{2} \, |\phi|^{2} \rangle$ in a perturbative manner. From that perspective, we study a four-point function for which we expect that any divergency $\sim p^{0}$ goes like $\ln\Lambda$ and the $\Lambda^{2}$ dependence is shifted to the terms ${\sim}p^{2}$.

However, whether this behavior will persist beyond our one-loop approximation requires a full two-loop analysis. 
Furthermore note that even if the $\Lambda^{2}$ dependence would cancel to all orders, it is not guaranteed that the hierarchy problem will be solved in a strict sense. In order to do so, similar effects have to manifest in all physical quantities that are related to the electroweak scale which is orders of magnitude smaller than the Planck scale. Exploring this direction might lead to new insights. 
If other physical observables show similar properties, a rethinking of the hierarchy problem will be mandatory. Once the classical scale is set via processes of strict gauge-invariant quantities in such a scenario, it only receives quantum corrections that do not alter the scale by several orders of magnitude.

\subsection{Resummation}
In order to extract the mass and the width from a simple pole of the propagator, a proper summation of self-energy diagrams is required for any elementary field. As the FMS mechanism allows us to project some of the nonperturbative bound state information on perturbatively accessible information, an appropriate resummation of certain contributions is required as well. At first glance, one may be tempted to resum the entire one-loop structure or rather its finite part stemming from all orders of the FMS expansion given in Eqs.~\eqref{eq:Prop-Bound-Ren} and \eqref{eq:Prop-Bound}. This is motivated by the fact that we obtain the standard form which can be written as $\frac{\I}{p^{2}-\mh^{2}} \frac{f(p^{2})}{p^{2}-\mh^{2}}$ where $f(p^{2})$ is the expression in curly brackets in Eq.~\eqref{eq:Prop-Bound} plus appropriate renormalization constants. Thus, one could naively conclude that $\I/(p^{2}-\mh^{2}-f)$ is a proper approximation for the bound state propagator in analogy to the one-loop resummation of the elementary propagator. However, one should keep in mind that we have chosen this particular one-loop structure by hand via inserting unities of the form $1=(p^{2}-\mh^{2})^{n}/(p^{2}-\mh^{2})^{n}$ for reasons of convenience to explicitly show the gauge-parameter cancelation among the different terms of the FMS expansion at this particular loop order. Realizing the different structures of the three different terms, cf. Eqs.~\eqref{eq:FMSo0}, \eqref{eq:FMSo1}, and \eqref{eq:FMSo2} or their diagrammatic representation in Fig.~\ref{fig:diagrams-schematically}, it becomes evident that a summation of the plain one-loop structure of the bound state is not a valid approximation as we would not treat the operator insertions in a correct manner. For instance, the terms $\frac{\I}{p^{2}-\mh^{2}} \Big(\frac{f(p^{2})}{p^{2}-\mh^{2}}\Big)^{n}$ cannot appear at $(n\geq 2)$-loop order as exponentiating $f$ would imply $n$ to $2n$ composite operator insertions. The FMS expansion is a finite series in $\varphi/v$ for the scalar bound state operator $|\phi|^{2}$ as it contains only a finite amount of elementary scalar doublets and thus the number of potential operator insertions is bounded from above by definition which is two for the current Higgs bound state operator.

Alternatively, we could use the definitions of Eqs.~\eqref{eq:FMSo0}, \eqref{eq:FMSo1}, and \eqref{eq:FMSo2}, plug them into Eq.~\eqref{eq:Propagator} and use the approximation $\Sigma_{i} \to \Sigma_{i}^{1\rml}$. This would account for the $|\varphi|^{2}$ insertions properly and leads to a straightforward perturbative treatment of the bound state propagator. It is given by the resummed elementary propagator in leading order. Slight modifications due to the higher-order terms of the FMS expansion account for the more involved internal structure. Unfortunately, this comes with another grain of salt. Suppose we perform an $n$-loop approximation for the three contributions $\Sigma_{0}$, $\Sigma_{1}$, and $\Sigma_{2}$ in the gauge-fixed setup. Then the resummed bound state propagator is indeed gauge-parameter independent up to $n$-loop order, but it will now explicitly depend on $\xi$ as we do not take $(n+1)$-loop diagrams appropriately into account. The resummation is simply flawed by the fact that we consider only a subgroup of all possible diagrams with loop order $n+1$ or higher. Thus, the gauge-parameter cancelation mechanism is not at work at these higher orders and we introduce an artificial gauge dependence due to our approximation. 
 
To circumvent this technical problem, we propose to only resum those contributions that are explicitly $\xi$ independent, i.e., to neglect any loop contributions that include at least one internal Goldstone, ghost, or longitudinal weak gauge boson line. This procedure is justified by the fact that the loop diagrams containing modes with masses ${\sim}\sqrt{\xi}m_{\mathrm{W/Z}}^{2}$ have to cancel order by order in the perturbative expansion by construction. More precisely, we simply use the identity 
\begin{align}
 \langle (\phid\phi)(p) \, (\phid\phi)(-p) \rangle  & = \frac{\I}{p^{2}-\mh^{2}-\Sigma_{0}} \Big(v^{2} + 2v\Sigma_{1}  \Big)  
  + \Sigma_{2} \notag \\
 & = \frac{\I  }{p^{2}-\mh^{2}-\tilde\Sigma_{0}}\Big(v^{2} + 2v\tilde\Sigma_{1}  \Big)  + \tilde\Sigma_{2}
\label{eq:resum-Bound}
\end{align}
where a function with a tilde $\tilde\Sigma_{i}$ is given by $\Sigma_{i}$ minus $\xi$-dependent contributions as defined at the end of Sec.~\ref{sec:one-loop} for the one-loop approximations. 
Using now approximations for the functions $\tilde\Sigma_{i}$ instead of $\Sigma_{i}$ and plug them into the last line of Eq.~\eqref{eq:resum-Bound}, we obtain a $\xi$-independent resummed bound state propagator that accounts for $|\varphi|^{2}$ insertions in the correct way. 
Thus, a consistently resummed one-loop approximation for the bound state propagator reads,\footnote{We suppress the counterterms for better readability in the following. We would like to emphasize here that they have to be treated on the same footing. In case the renormalization constants are fixed via the gauge-dependent $n$-point functions, any $\xi$-dependent contribution of the finite part has to be neglected as well. Equivalently, we can choose a scheme that does not include the $\xi$-dependent (finite) parts in the fixing procedure, see the discussion in Sec.~\ref{sec:ren-bound}.}
\begin{align}
 &\langle (\phid\phi)(p) \, (\phid\phi)(-p) \rangle  \notag \\
 &\quad= \frac{\I}{p^{2}-\mh^{2}-\tilde\Sigma_{0}^{1\rml}} \Big(v^{2} + 2v\tilde\Sigma_{1}^{1\rml} + (\tilde\Sigma_{1}^{1\rml})^{2} \Big)  
  + \tilde\Sigma_{2}^{1\rml}.
\label{eq:resum-Bound-1loop}
\end{align}
Here, we have treated the resummation of diagrams contributing to $\Sigma_{2}$ on the same footing as for $\Sigma_{0}$ and $\Sigma_{1}$, i.e., we included the sum of any diagram containing an arbitrary number of one-loop diagrams connected via tree propagators. This implies that we have $\tilde\Sigma_{2} = \tilde\Sigma_{2}^{1\rml} + \frac{\I}{p^{2}-\mh^{2}-\tilde\Sigma_{0}^{1\rml}}(\tilde\Sigma_{1}^{1\rml})^{2}$ with $\tilde\Sigma_{2}^{1\rml}\equiv \tilde{\Gamma}^{(0,2)1\rml}$. Equivalently, one could impose a resummation scheme that only resums elementary Higgs self-energy contributions up to a given order and truncate the functions coming from higher orders of the FMS expansion in the numerator at the same loop order. Implementing this resummation strategy implies that we would neglect the term $(\tilde\Sigma_{1}^{1\rml})^{2}$ in Eq.~\eqref{eq:resum-Bound-1loop} as we would assign it to the two-loop approximation.

These two strategies provide a useful treatment of the bound state propagator that are valid in any renormalization scheme. For particular schemes, further options are feasible, e.g., within the on-shell or complex mass scheme where renormalized parameters are identified with measurable quantities. For instance, we could consider a resummation that treats all terms in the numerator at the same loop order, i.e., 
\begin{align}
&\frac{\I \Big[v^{2}\frac{\tilde\Sigma_{0}^{1\rml}}{p^{2}-\mh^{2}} + 2v\tilde\Sigma_{1}^{1\rml} + \mathcal{O}(\hbar^{2}) \Big]}{p^{2}-\mh^{2}-\tilde\Sigma_{0}^{1\rml}} + \tilde\Sigma_{2}^{1\rml}  \notag \\
&\quad= \I\frac{v^{2}\frac{\tilde\Sigma_{0}^{1\rml}}{p^{2}-\mh^{2}} + 2v\tilde\Sigma_{1}^{1\rml} + (p^{2}-\mh^{2})\tilde\Sigma_{2}^{1\rml}  + \mathcal{O}(\hbar^{2})}{p^{2}-\mh^{2}-\tilde\Sigma_{0}^{1\rml}} 
\label{eq:resum-Bound-1loop-2}
\end{align} 
at the one-loop level and more sophisticated expressions at higher loop orders. It is clear that such a strategy can only be valid in those particular schemes where $\mh$ is identified with the Higgs mass as it would otherwise introduce additional RG scale dependent fake poles, e.g., within the $\overline{\mathrm{MS}}$ scheme.

At our current level of approximation, it does not matter which of the different options we choose as the modifications are negligible for our present purpose. We have tested this by comparing the extracted spectral functions of the different resummation options whose general properties are discussed below. Nevertheless, depending on the actual objectives and the requirements on the precision of the calculation for more sophisticated investigations, some of the proposed strategies on how to deal with the diagrams of the higher-order FMS terms might be more useful than others.\footnote{A word of caution. We have outlined different resummation strategies which follow from the FMS expansion in a  natural way. Following these strategies in a consistent manner is important to obtain reliable results as slight deviations from these expansions can give inadequate convergence properties.}
For the present case, we merely used them as a nontrivial test to evaluate our one-loop approximation. The different options only give different results for the resummed propagator if the self-energy functions $\tilde{\Sigma}_{i}$ are calculated within a given approximation. In case $\tilde{\Sigma}_{i}^{\mathrm{approx.}} \to \tilde{\Sigma}$, this seeming ambiguity will vanish and the properly resummed propagator is uniquely defined.

Before we analyze the physical information encoded in the resummed bound state propagator, we have to discuss a further issue. 
The finite part of $\tilde\Sigma_{0}^{1\rml}$ behaves as $p^{4}\ln p^{2}$ for large $p^{2}$. Thus naively summing the one-loop approximation of the self energy, $\tilde\Sigma_{0}^{1\rml}/(p^{2}-\mh^{2})$ becomes nonconvergent for sufficiently large $p^{2}$. This problem already appears at the level of the elementary Higgs propagator, unless we choose $\xi=1$. Due to the interplay of the different modes, however, $|\Sigma_{0}^{1\rml}/(p^{2}-\mh^{2})|<1$ over a sufficiently large range of the external momentum $p^{2}$ if the gauge-fixing parameter is sufficiently small. For $\xi< 40$ this requirement is fulfilled for any energy range accessible by present and near-future colliders.  For the bound state propagator, we have $|\tilde\Sigma_{0}^{1\rml}/(p^{2}-\mh^{2})|<1$ only up to $\approx 1$ TeV. Thus, the naive resummation of terms ${\sim} \frac{p^{4}\ln p^{2}}{p^{2}-\mh^{2}}$ is only meaningful below this scale. 

In order to take this into account, we could either restrict our study only to this energy range or simply exclude this particular contribution from the resummation and keep it at an appropriate loop level.\footnote{Technically, one expects similar issues from terms being ${\sim} \frac{p^{2}\ln p^{2}}{p^{2}-\mh^{2}}$ for large $p^{2}$. On a practical level, however, these terms do not spoil resummation at any relevant physics scale as they become $\mathcal{O}(1)$ far beyond the Planck scale.} The latter option can straightforwardly be implemented at the one-loop level if the on-shell/complex mass scheme is used. In this case, we are able to use 
\begin{align}
 \langle (\phid\phi)(p) \, (\phid\phi)(-p) \rangle
 & = \frac{\I \Big[v^{2} + 2v\tilde\Sigma_{1}^{1\rml} + (\tilde\Sigma_{1}^{1\rml})^{2} \Big]}{p^{2}-\mh^{2}-\hat{\tilde\Sigma}_{0}^{1\rml}}   
   \notag \\
  &\quad+ \tilde\Sigma_{2}^{1\rml} + v^{2}\frac{\I \tilde\Sigma_{0,p^{4}}^{1\rml}}{(p^{2}-\mh^{2})^{2}},
\label{eq:resum-Bound-1loop3}
\end{align}
where $\tilde\Sigma_{0,p^{4}}^{1\rml} = -\frac{p^{4}}{2v^{2}}\big[2\loopB(\mW^{2}) + \loopB(\mZ^{2})\big]$ and $\hat{\tilde\Sigma}_{0}^{1\rml} = \tilde\Sigma_{0}^{1\rml} - \tilde\Sigma_{0,p^{4}}^{1\rml}$. For any other scheme, we would spoil the calculation by introducing additional poles. This might be cured by introducing Heaviside step functions or smooth regulator functions such that the resummation of the problematic terms can be switched on and off at appropriate momentum scales above the pole but below the energy range where the resummation breaks down. Alternatively, one may follow the strategy of Ref.~\cite{Dudal:2019pyg} where a partial resummation is proposed. Using the identity $p^{4} = (p^{2}-\mh^{2})^{2} +2\mh^{2}p^{2} - \mh^{4}$, we can split the one-loop contribution $p^{4}\ln(p^{2})/(p^{2}-\mh^{2})^{2}$ into a pure logarithm, which will not be resummed and only features the common branch cuts as well as a remaining term that will be included into the resummation. Within our accuracy, we did not find any significant numerical deviation between this method and Eq.~\eqref{eq:resum-Bound-1loop3} for the on-shell or complex mass scheme. Nevertheless, we have to state that the (one-)loop approximation is not able to capture the external momentum dependence properly at large energies. In order to gain a glimpse of the potential behavior slightly above $1$ TeV, we will use Eq.~\eqref{eq:resum-Bound-1loop3}. However, we emphasize that improvements are necessary to make a conclusive statement at TeV scales, e.g., via Pad\'{e} approximations or resummation techniques that include large logarithms. As we are mainly interested in the properties near the pole, where the summation converges as substantiated by the fact that the methods of Ref.~\cite{Dudal:2019pyg} and Eq.~\eqref{eq:resum-Bound-1loop3} give same results, and not the high-energy momentum tail of the propagator, we will use our simple summation in the following.

\subsection{Mass, width, and spectral function}
After this preparatory work, we are now able to discuss the phenomenological consequences of the perturbative treatment of the Higgs bound state via the FMS formalism. First of all, we find that the mass of the bound state $\phid\phi$ is identical to the mass of the elementary Higgs field $h$ within gauges that allow for a nonvanishing VEV. This is explicitly demonstrated in terms of a one-loop approximation, by comparing the pole structure of Eq.~\eqref{eq:resum-Bound-1loop} with its elementary counterpart $\langle h\,h\rangle = \I/(p^{2}-\mh^{2}-\Sigma_{0}^{1\rml})$. Most importantly, this feature will also be present at any loop approximation as can be seen by the general structure of the FMS expanded but properly loop-order resummed bound state propagator, see Eq.~\eqref{eq:resum-Bound}. Within a perturbative treatment, the higher order terms of the FMS expansion do not alter the mass and width extracted from the position of the pole which is consistent with the Nielsen identities. Furthermore, they will also most likely not introduce any new pole structure that is not already contained within the elementary propagator. 

\begin{figure}
\centering
\includegraphics[width=0.35\textwidth]{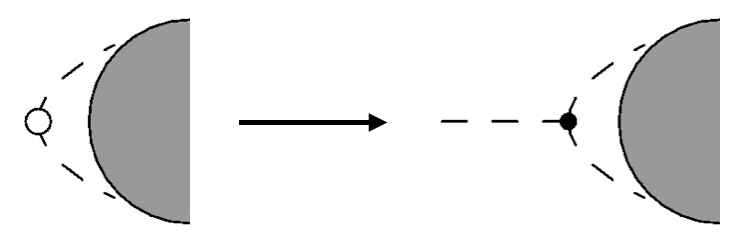}
\caption{Diagrammatic sketch that all diagrams contributing to next-to-leading and higher orders of the FMS expansion can be converted into a subgroup of all diagrams that contribute to the leading order term, i.e., the propagator of the elementary field.}
\label{fig:converted}
\end{figure}

The latter fact can be made transparent by recognizing that any diagram of $\langle h\, |\varphi|^{2} \rangle$ and $\langle |\varphi|^{2}\, |\varphi|^{2} \rangle$ can be converted into a diagram contributing to $\langle h\, h \rangle$ by replacing the operator insertion via a $h\varphid\varphi$ vertex and an external $h$ line, see Fig.~\ref{fig:converted}. This has important consequences. It shows that even if some internal excitation of the bound state exist due to the complex interplay of the constituents, it cannot be addressed in a gauge with nonvanishing $v$ unless it is encoded in the elementary Higgs propagator as all appearing structures are constrained by the FMS expansion. In general, this argument can also be used for all other $\SU(2)$ gauge-invariant descriptions of standard model particles. Of course this is not a weakness of the FMS formalism as it still describes all occurring structures in a gauge-invariant manner. It is rather that the standard $R_{\xi}$ gauges would not be able to properly reflect the entire physical information of the system. They would describe the ground state but additional information might be missing. Nonetheless, nonperturbative lattice simulations confirm that at least within the $\SU(2)$-Higgs subsector the perturbative description seems to capture all relevant information \cite{Wurtz:2013ova,Maas:2014pba}. In particular the one-loop approximation is decent for the lattice results, see the next Sec.~\ref{sec:lattice}.

Although the pole position of the bound state and the elementary field is identical, we do obtain phenomenological implications as cross sections get altered due to the higher-order terms of the FMS expansion. For the sake of illustration, let us consider the square of the transition amplitude of 1-to-1 scattering. 
We have 
\begin{align}
 |\mathcal{M}_{\mathrm{h} \to \mathrm{h}}|^{2} = \frac{1}{|p^{2}-\mh^{2}-\Sigma_{0}|^{2}}
\end{align}
and 
\begin{align}
 |\mathcal{M}_{\phid\phi \to \phid\phi}|^{2} &= \frac{1+\frac{4}{v}\re \tilde\Sigma_{1} + \frac{4}{v^{2}}|\tilde\Sigma_{1}|^{2}}{|p^{2}-\mh^{2}-\tilde\Sigma_{0}|^{2}} + |\tilde\Sigma_{2}|^{2}  \notag\\
 &\quad+ 2\re \bigg(\frac{1+\frac{2}{v}\tilde\Sigma_{1}}{p^{2}-\mh^{2}-\tilde\Sigma_{0}}\tilde\Sigma_{2} \bigg)
\end{align}
for the elementary field and the bound state transition, respectively. Note that we used an appropriate normalization of the bound state propagator, $\langle |\phi|^{2}\, |\phi|^{2} \rangle \to \frac{1}{v^{2}}\langle |\phi|^{2}\, |\phi|^{2} \rangle$, which is equivalent to considering the rescaled operator $\frac{1}{v}\phid\phi$, to provide a comparison with the elementary field transition amplitude.

\begin{figure*}
\centering
\hfill
\includegraphics[width=0.475\textwidth]{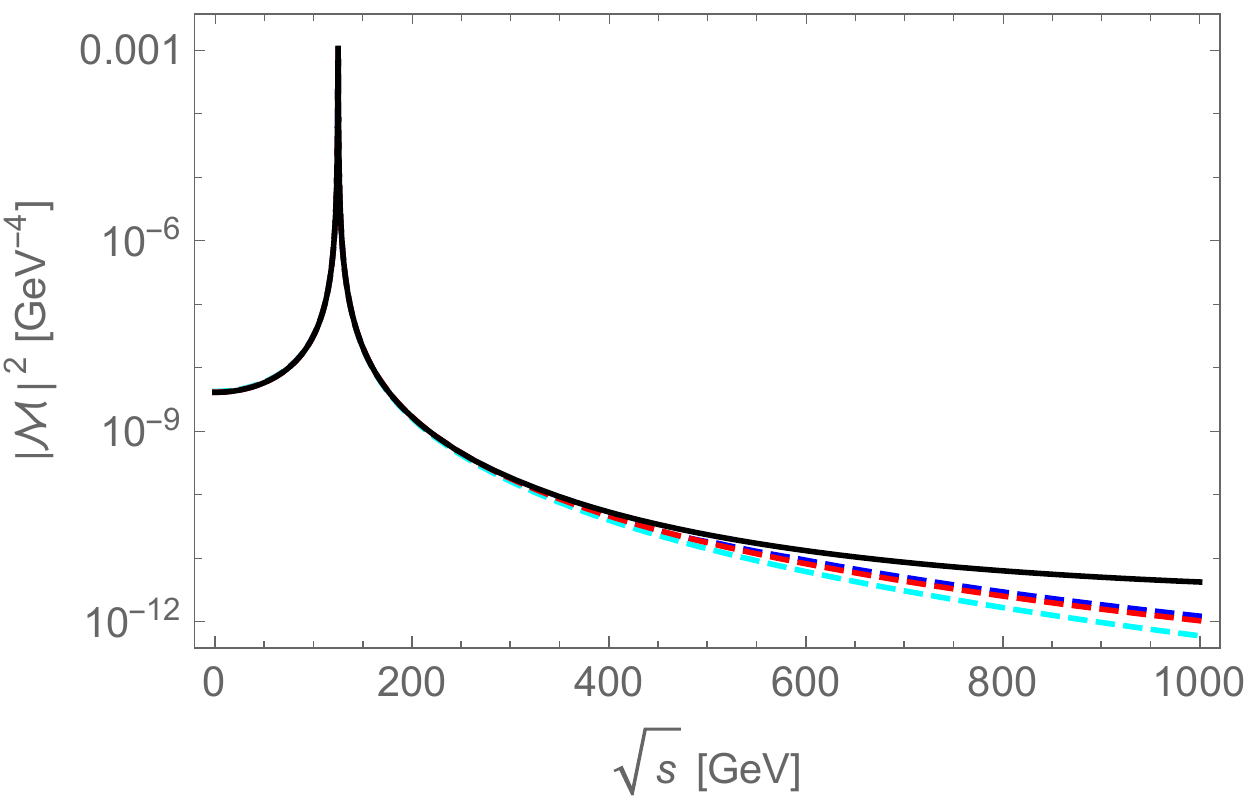}
\hfill
\includegraphics[width=0.475\textwidth]{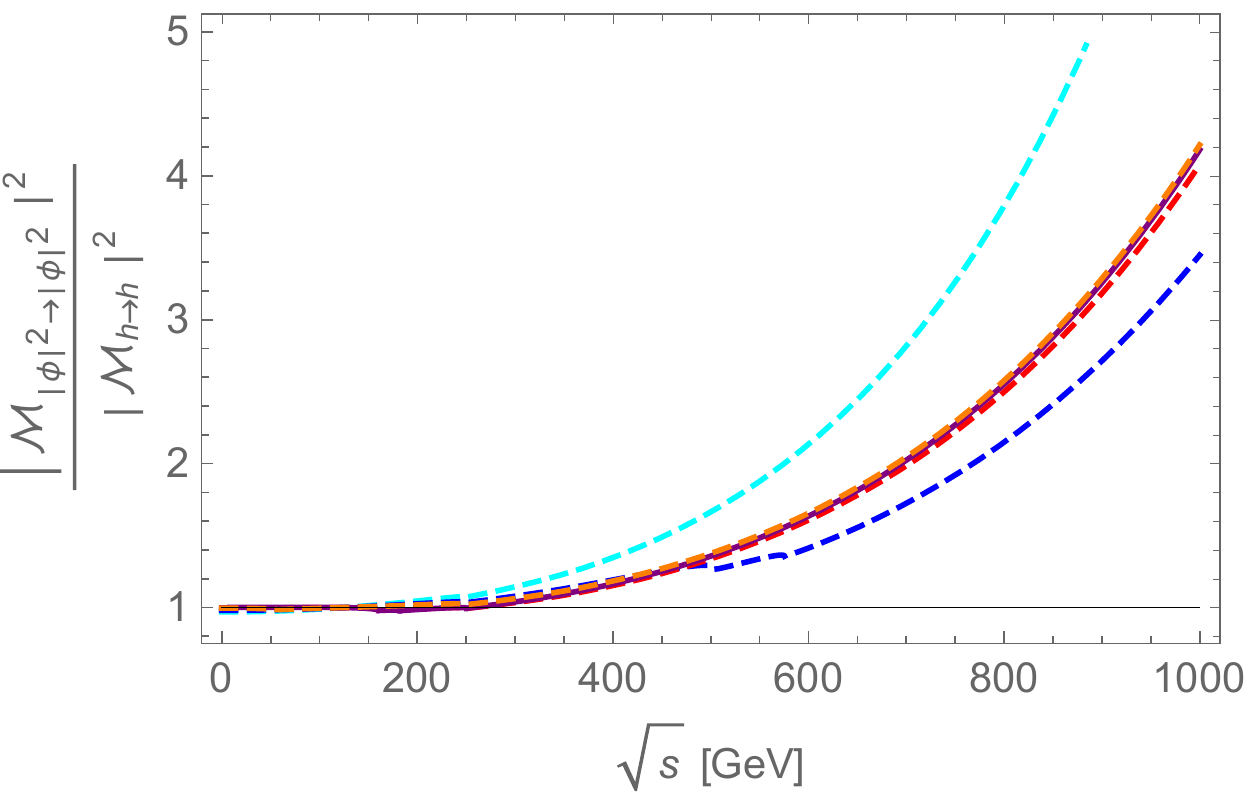}
\hfill
\caption{Left panel: transition amplitude of ``1-1 scattering'' for the bound state (black solid line) and elementary Higgs (dashed lines). The red dashed line denotes the elementary transition amplitude for $\xi=1$, while the blue line denotes $\xi=10$ and the cyan line $\xi=100$. Right panel: ratio of the bound state and elementary field transition amplitude. We also included the results for the pinch technique propagator (solid purple line) and $\xi=0.1$ (orange dashed line). These two lines cannot be distinguished from the red curve in the left panel by the eye.}
\label{fig:scattering}
\end{figure*}

The transition amplitudes are depicted in the one-loop approximation for $s=p^{2}$ in the left panel of Fig.~\ref{fig:scattering}. The black solid curve shows $|\mathcal{M}_{|\phi|^{2} \to |\phi|^{2}}|^{2}$ while dashed lines depict $|\mathcal{M}_{\mathrm{h} \to \mathrm{h}}|^{2}$ for $\xi = 1$ (red), $\xi = 10$ (blue), $\xi = 100$ (cyan). For $\xi < 1$ as well as for the pinch technique propagator we do not observe any deviation from the red dashed curve that is visible by eye. In the right panel, we plotted the ratio of the bound state operator with the elementary field for different values of the gauge-fixing parameter. For illustration, we also included $\xi=0.1$ (orange) as well as the ratio of the bound state amplitude and the amplitude obtained from the pinch technique propagator (purple solid line). In the small external momentum regime, we find good agreement of all different functions. At $\sqrt{s} = 125$ GeV all transition amplitudes coincide as they should due to the $\xi$ independence of the pole of the propagator. However, note that we technically get a tiny displacement of the location of the maximum of the peak of the relativistic Breit-Wigner distribution and its width away from the pole position for the transition amplitude of the bound state operator. This shift is caused by the modified structure in the numerator due to the higher-order contributions of the FMS expansion. On a practical level, however, we do not obtain an observable alteration of the peak for standard model parameters. 

Away from the narrow peak, we have deviations of $\mathcal{O}(1\%)$ for $\xi=1$ below the two-Higgs threshold. The deviation increases with increasing $\xi$ in this momentum region, e.g., up to $7\%$ at the two-Higgs threshold for $\xi=100$. The transition amplitude computed from the pinch technique Higgs propagator stays close to the bound-state transition amplitude below any bosonic threshold and gets $\approx 2\%$ suppressed at the $W$ and $Z$ thresholds. In the energy range up to $1$ TeV, we obtain an enhancement of the bound-state transition amplitude compared to $\mathcal{M}_{\mathrm{h}\to\mathrm{h}}$. We find that this enhancement first decreases with increasing $\xi$ at fixed $s$ up to $\xi \approx 17$. For larger $\xi$ the enhancement increases. The result obtained from the pinch technique is close to the result of the Feynman--t' Hooft gauge. In the energy region of $600$ to $800$ GeV, we observe an enhancement of the transition amplitude of the bound state Higgs compared to the pinch technique propagator by a factor of $2$.

As the results of the bound state computation shows deviations from the results conventionally obtained by the pinch technique, processes that involve an intermediate Higgs would be a possible test ground to examine the phenomenological impact of the FMS formulation. However, note that our current results need further improvements as we purely considered propagatorlike diagrams so far but neglected any contributions from vertex or box diagrams which are important within the pinch technique treatment.

\begin{figure}
\centering
\includegraphics[width=0.45\textwidth]{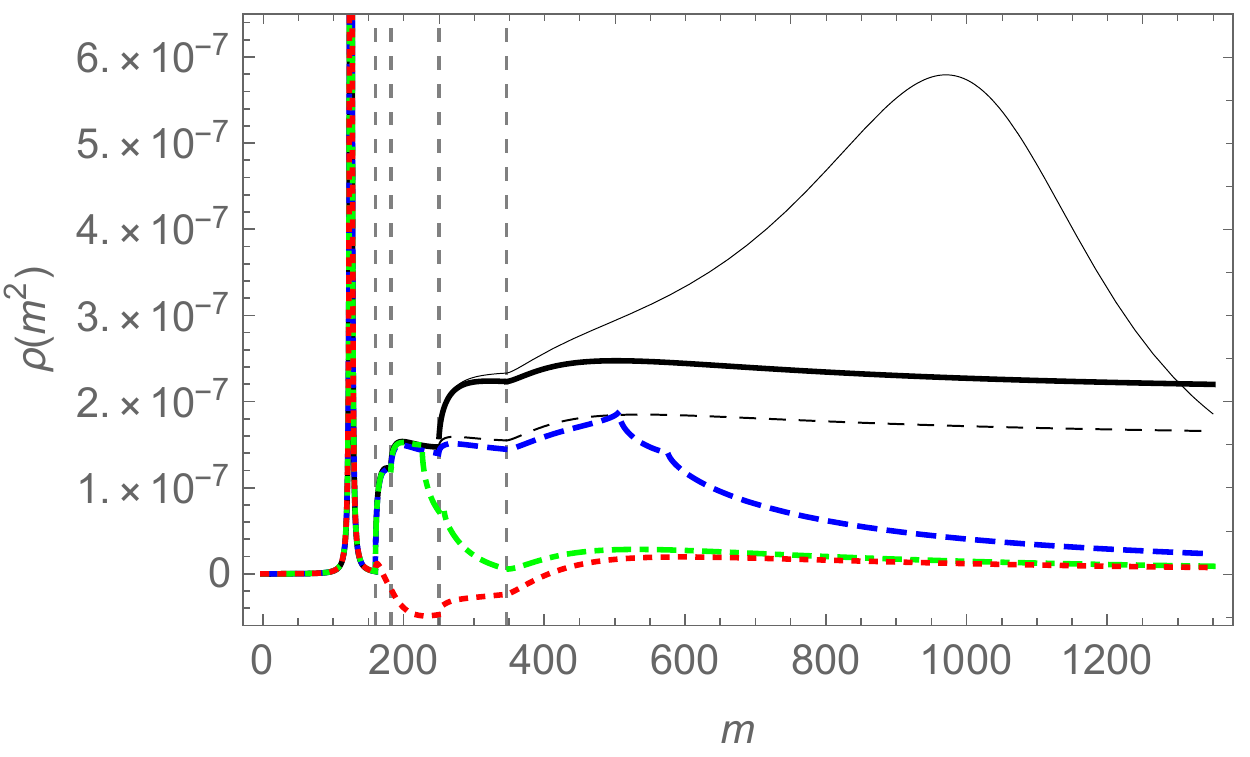}
\caption{Spectral density of the elementary Higgs field for different values of the gauge-fixing parameter $\xi$ and of the bound state operator $\phid\phi$ (black solid line). We depict the spectral function for $\xi=1$ (red dotted line), $\xi=2$ (green dash-dotted line), and $\xi=10$ (blue dashed). The vertical gray dashed lines indicate the mass thresholds at $2\mW$, $2\mZ$, $2\mh$, and $2m_{\mathrm{top}}$ from left to right. Further, we depicted the spectral function obtained from the bound state propagator but all nontrivial effects from $\tilde\Sigma_{1}$ and $\tilde\Sigma_{2}$ ignored as a black dashed line. The thin black solid line illustrates the result if we would had included terms $\sim p^{4}\ln p^{2}$ into the resummation.}
\label{fig:spectralBound}
\end{figure}

Finally, let us consider the spectral function of the bound state propagator which is depicted in Fig.~\ref{fig:spectralBound}. The most important result is the fact that it is non-negative. This is a basic requirement for a physical interpretation which is not possible for the spectral function of the elementary field. For comparison, we also plotted as thin black dashed line the spectral function from a toy propagator that contains the $\xi$-independent Higgs self-energy $\tilde\Sigma_{0}$ but all other effects coming from the higher-order terms of the FMS expansion ignored. Comparing the thin black dashed and thick solid line, we find a relevant enhancement of the spectral function due to the nontrivial internal properties of the bound state operator described by the higher-order FMS terms. The enhancement starts at the two-Higgs threshold. This is expected as we only obtain in this energy range a significant alteration of the imaginary part of the propagator at the one-loop level due to the higher-order terms. Additionally, we plotted as thin black solid curve the spectral function extracted from a bound state propagator where the $p^{4}\ln p^{2}$ terms of the self-energy $\tilde\Sigma_{0}^{1\rml}$ are resummed as well. This would provide a further enhancement around $1$ TeV and above this scale a fast decrease. Nevertheless, we should not trust the one-loop approximation in this regime. Higher-order computations are needed to make a conclusive statement in this energy range.

\section{Latttice}
\label{sec:lattice}
So far, we computed the bound state properties purely by perturbative methods. The fact that this is possible is a particularity of gauge theories with a BEH mechanism due to the FMS relation in certain classes of gauges. In the following, we will test whether a perturbative treatment of the FMS expansion is able to capture most of the relevant information of the Higgs particle in the weak coupling regime or nonperturbative bound state effects might occur that lead to further phenomenological implications.

In order to examine the validity of the FMS expansion, we compare our one-loop results with lattice simulations. Therefore, we switch to Euclidean signature and focus the analysis to the pure bosonic sector. The latter restriction is due to the present unavailability of an efficient algorithm to simulate gauged Weyl fermions. Furthermore, we take the limit of coinciding $W$ and $Z$ boson masses, i.e., we investigate a non-Abelian $\SU(2)$-Higgs model as this theory is extensively studied on the lattice, e.g., see \cite{Maas:2012tj,Maas:2013aia,Maas:2014pba,Maas:2018ska} or \cite{Maas:2017wzi} for a review.

From a conceptual viewpoint it would be most convenient for the sake of comparison to impose renormalization conditions at vanishing Euclidean momentum for our one-loop analysis and the lattice data. However, finite volume effects are usually expected and have to be treated carefully for low momenta within the lattice setup. For the analysis of the elementary Higgs propagator it was thus chosen to impose renormalization conditions at $|p|=\mh$ such that the lattice propagator is fixed to the tree-level one at this scale \cite{Maas:2010nc}. As we extend the analysis to the bound state propagator and examine various parameter sets, we use a different strategy to avoid potential issues coming from interpolation and discretization artifacts. 
For each propagator, we choose the third lowest momentum obtained from the lattice results under consideration as a reference point. Then, we use the freedom of the finite parts of the mass and wave function renormalization constants to impose renormalization conditions such that the lattice propagator coincides with the momentum-space propagator obtained from the loop expansion at this point. By using this strategy, we minimize a potential contamination of the intermediate and large momentum regime from finite volume effects. At the same time, we avoid a contamination of the low and intermediate momentum regime  from discretization artifacts above the lattice cutoff.

\begin{figure*}
\centering
\includegraphics[width=0.45\textwidth]{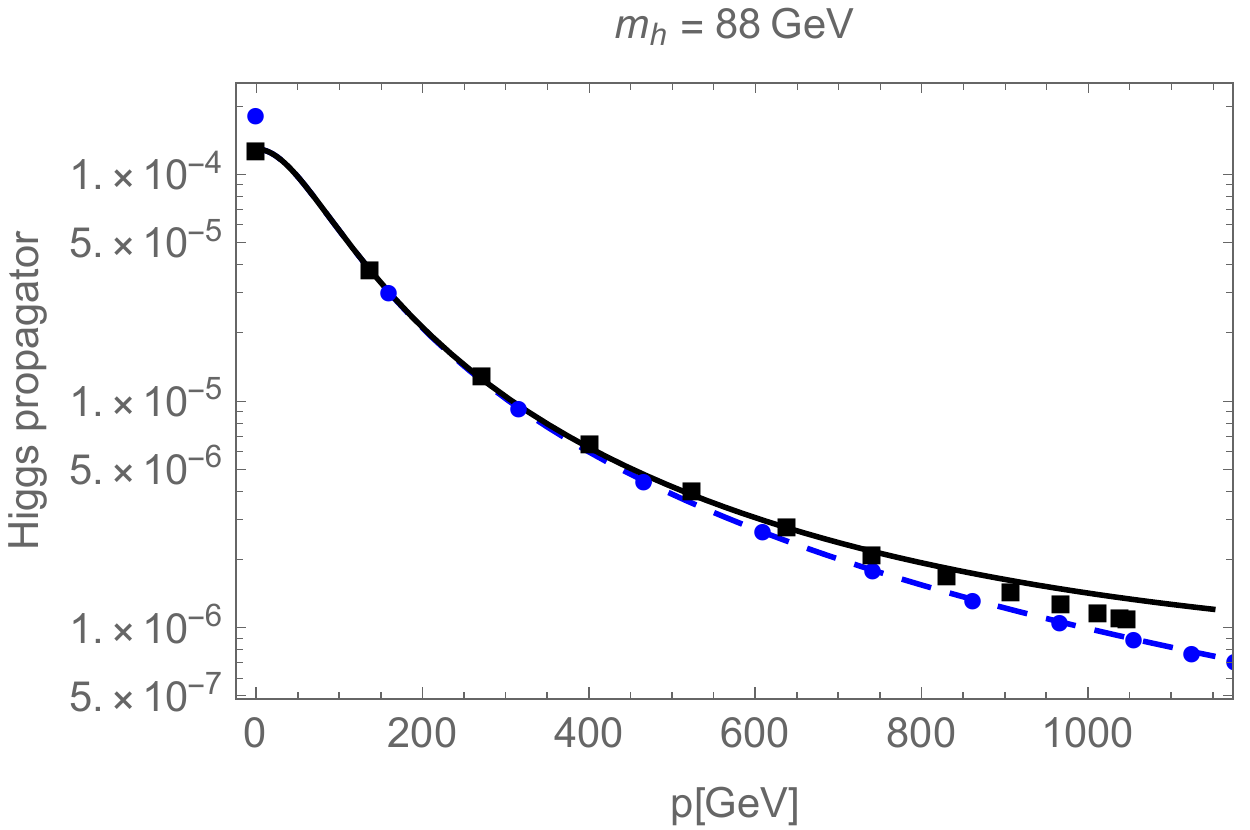} \hfill
\includegraphics[width=0.45\textwidth]{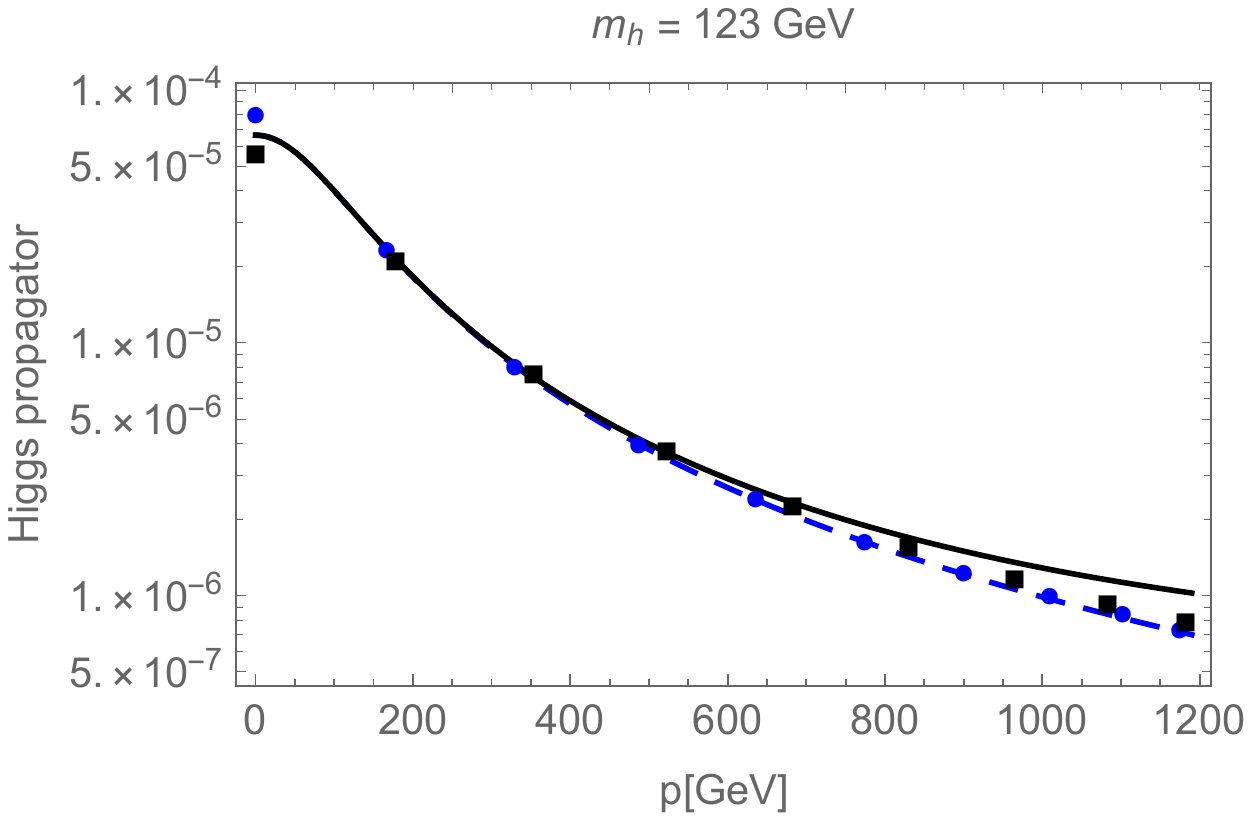}\hfill
\includegraphics[width=0.45\textwidth]{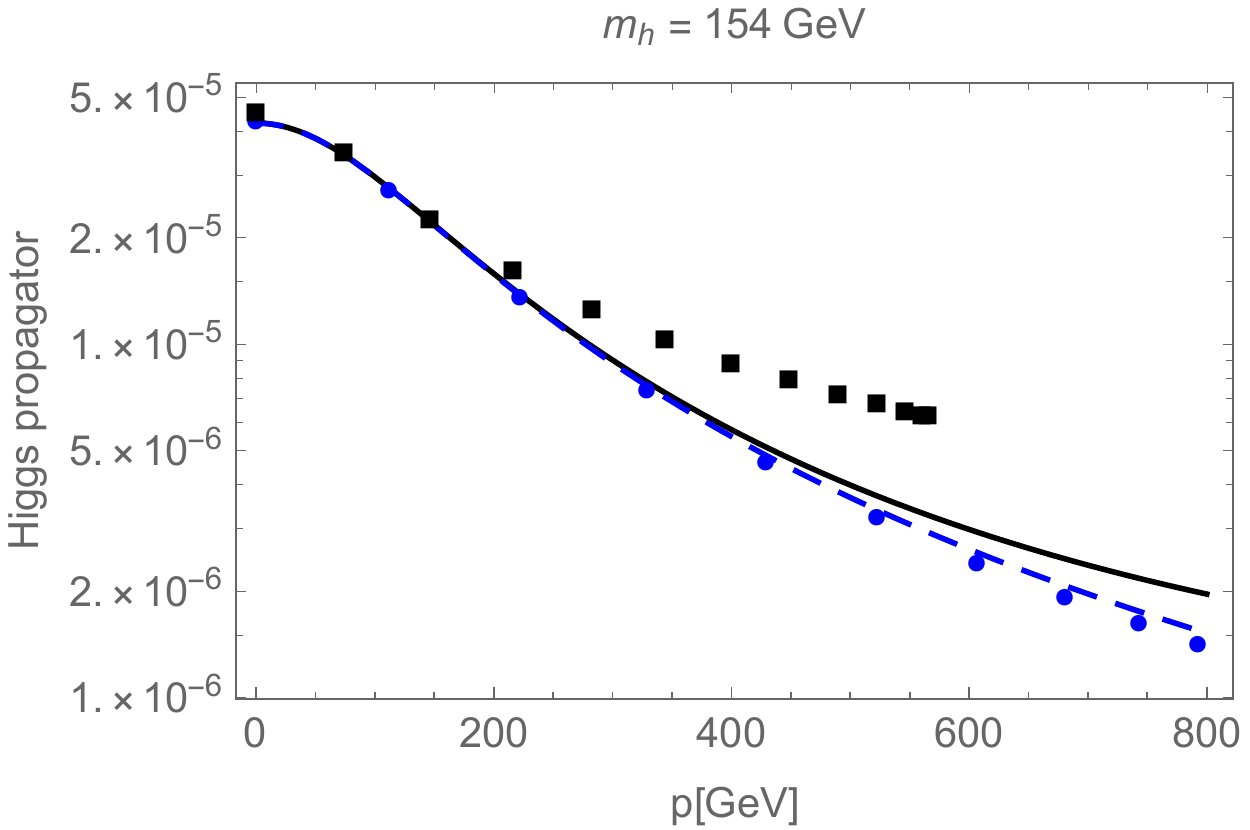}
\caption{Comparison between the momentum-space propagator extracted from nonperturbative lattice simulations and the Euclidean one-loop approximation for different masses $\mh$ within an $\SU(2)$-Higgs model. Blue circles and black squares represent lattice results \cite{Maas:2013aia} for the elementary and bound state propagator, respectively. Accordingly, blue dashed lines show the elementary one-loop propagator while black solid lines show the bound state propagator.}
\label{fig:PropLat}
\end{figure*}

In Fig.~\ref{fig:PropLat}, we compare lattice and analytical one-loop results for the elementary (blue) and bound state (black) propagator for different ratios of the scalar and vector boson mass.\footnote{While gauge-fixing on the lattice faces generically the Gribov-Singer ambiguity, we do not expect that this will affect the results here. On the one hand, it has been argued that Gribov copies become quantitatively irrelevant in the BEH case \cite{Lenz:2000zt}, which is corroborated by the fact that they seem to be essentially absent in lattice simulations in this case, and even if present, they do not affect the Higgs propagator measurably \cite{Maas:2010nc}. In addition, even for the gluon, whose propagtor in Yang-Mills theory is strongly affected by Gribov copies, its analytic properties as encoded in its Schwinger function are essentially unaffected \cite{Maas:2017csm}. This is likely due to the case that Gribov copies live on a very different length scale than those relevant for dynamics \cite{Heinzl:2008bu}.} 
For all results, we use $\mW = 80$ GeV. Statistical error bars for the lattice data are suppressed as they are at most at the one percent level and hardly visible by eye. The three different plots show the Higgs propagator for $\mh=88$ GeV, $\mh=123$ GeV, and $\mh=154$ GeV. 
Apart from finite volume effects, we obtain decent results for the elementary propagator. For the two light Higgs masses, the analytical and lattice results differ by $\mathcal{O}(1\%)$. In the large momentum regime, $p\gtrsim 1000$ GeV, we observe differences of 5\% but this momentum range is already above the lattice cutoff scale ($\approx 620$ GeV for both datasets). Thus, these differences are likely affected by discretization artifacts \cite{Maas:2013aia,Maas:2016edk,Maas:2018ska}. At $p=0$, we get large differences of ${\sim}25\%$ ($\mh=123$ GeV) and ${\sim}40\%$ ($\mh=88$ GeV). However, on top of the finite-volume artifacts \cite{Maas:2013aia} we used the central values for the masses obtained from a spectroscopic analysis in the scalar channel which have similar uncertainties due to finite volume effects, $\mh=123 \pm 19$ GeV and $\mh=88 \pm 10$ GeV, respectively. Taking these aspects into account, the $p=0$ regime is within this uncertainty in the $\mh=123$ GeV case and at least close to the lower value of the mass interval for $\mh= 88$ GeV. We obtain a similar picture for the bound state propagators. We find finite volume effects of $\mathcal{O}(10\%)$ at $p=0$. At finite Euclidean momentum, but below the lattice cutoff, we find deviations of a few percent between the lattice and analytical results. In the large momentum regime, this deviation increases to $10\%$ ($p{\sim}800$ GeV) and $25$--$40\%$ ($p{\sim}1$--$1.2$ TeV). Especially for the bound state propagators, this is likely due to discretization artifacts, as well as mixing with scattering states \cite{Maas:2018ska,Maas:2017wzi}.

For the largest Higgs mass $\mh=154 \pm 5$ GeV, we obtain similar results for the elementary propagator as for the two lighter Higgs masses. At $p=0$, we have only a slight deviation of $2\%$ between the lattice and analytical results which is compatible with the uncertainty of the scalar mass due to finite volume effects. In the intermediate momentum regime, we have deviations of $\mathcal{O}(1\%)$ but observe larger deviations of ${\sim}4$--$10\%$ above the lattice cutoff at $429$ GeV. For the bound state propagator, we obtain good agreement between both methods sufficiently below the lattice cutoff. At $p\gtrsim 300$ GeV, however, we find larger deviations. This may be caused by several facts. First, the scalar self-coupling is stronger than in the other two cases such that a simple one-loop approximation of the FMS terms might not reflect all relevant bound state information. Second, the $\mh=154$ GeV is close to the two-$W$ threshold at $160$ GeV. Thus, we might see stronger mixing effects with scattering states than in the previous two cases, which alter the high-momentum tail substantially \cite{Maas:2018ska}.

\section{Summary and Conclusion}

We investigated a strict gauge-invariant definition of the standard model Higgs boson resonance given by the FMS framework. Usually the properties of correlation functions of the elementary Higgs field entering the Lagrangian are considered to make contact with collider experiments. The main advantage of this approach is the comparatively straightforward computation of these quantities with perturbative methods. However, these correlation functions depend explicitly on the chosen gauge and the extraction of gauge-parameter invariant information can be challenging. By contrast, the main advantage of the FMS approach is given by the explicitly gauge-invariant formulation of all Green's functions such that the extraction of gauge-invariant physical quantities is (almost) trivial. This property is based on the fact that composite local bound state operators are considered instead of the elementary fields. The price to be paid for explicit gauge-invariance from the beginning is a more involved treatment of in principle nonperturbative objects. 

As shown by the original FMS work \cite{Frohlich:1980gj,Frohlich:1981yi}, some properties of the gauge-invariant bound states can be directly computed  from the $n$-point functions of the elementary fields if standard gauge-fixing procedures are used that allow for a nonvanishing vacuum expectation value of the scalar doublet. In particular, the bound state operator $\phid\phi$ can be mapped in leading order on the usual Higgs fluctuation mode $h$ such that the main phenomenological consequences of both approaches coincide.

For the first time, we go beyond this and address the impact of the higher order terms of the FMS expansion for the standard model Higgs in detail. As a main result, we were able to show that the pole structure of the elementary propagator coincides to all orders in the perturbative expansion with the pole structure of the bound state propagator which is given by all terms on the right-hand side of the finite FMS expansion. Thus, the mass and decay width of the gauge-invariant bound state operator is indeed well described by its elementary counterpart. This also gives a coherent picture with respect to the Nielsen identities which demonstrate the gauge-parameter invariance of the Higgs mass extracted from the elementary propagator in particular classes of gauges. Albeit the description of on-shell properties are the same within our investigations, we identified potential deviations regarding the Higgs off-shell properties if a virtual bound state Higgs is created in some process. Considering only propagatorlike diagrams, we find an enhancement of the transition amplitude of the bound state Higgs compared to the pinch technique propagator by a factor of $2$ in the energy range of $600$ to $800$ GeV. This might be an interesting starting ground to test the FMS implications besides other suggestions \cite{Egger:2017tkd,Maas:2018ska,Fernbach:2020tpa}. 

Additionally, the formulation and interpretation of the Higgs resonance in terms of the FMS bound state operator has further advantages. From a conceptual point of view, various field-theoretical issues are circumvented. For instance, no notion of a spontaneously broken electroweak gauge symmetry is required. Furthermore, the Lehmann-K\"all\`{e}n spectral density of the elementary Higgs field is plagued by positivity violations that spoil any physical interpretation of the elementary Higgs. This problem does not occur at the bound state level whose spectral density is gauge invariant, does not possess unphysical thresholds, and is strictly nonnegative. 

Further, we discussed the renormalization of the bound state propagator in detail. 
A peculiar situation appeared with respect to the bound state mass renormalization. In contrast to the elementary Higgs mass renormalization, the bound state mass term depends only logarithmically on the scale of new physics if the standard model is viewed from an effective field theory point of view. Thus, no fine-tuning of the scalar mass parameter seems to be required to obtain a Higgs mass orders of magnitude smaller than the Planck scale. As to whether this feature extends beyond the one-loop approximation is an open problem. Interestingly, this finding is curiously reminiscent of observations made in lattice simulations of the bosonic sector \cite{Maas:2017wzi}. However, even if this might be the case for the standard model Higgs mass, the hierarchy problem might still manifest itself in other physical quantities which are related to the electroweak scale. Nonetheless, further investigations into this direction might lead to novel insights into the underlying standard model cutoff scale.

In order to obtain a nontrivial test for the validity of the perturbative treatment of the FMS expansion, we compared our one-loop approximations for the elementary Higgs field propagator as well as for the bound state propagator against nonperturbative lattice simulations within the weak $\SU(2)$-Higgs subsector of the standard model. We gained decent results for Higgs masses sufficiently below the two-$W$ threshold scale where slight deviations could be traced back to finite volume or discretization artifacts. In case the Higgs mass comes close to the two-$W$ threshold, the deviations of the lattice data and the analytical bound state propagator increased. This might be caused by the impact of mixing with scattering states on the lattice or an insufficient description of the propagator by the one-loop approximation at large momenta. Nevertheless, in case of a physical Higgs-to-$W$ mass ratio, we find a good agreement between lattice and analytical results. Therefore, a perturbative description of all terms appearing within the FMS expansion seems to be indeed a useful approximation for the standard model.

\section*{Acknowledgement}
We are grateful to Holger Gies for valuable discussions. R. S. acknowledges support by the DFG under Grant No. SO1777/1-1.

\bibliography{bibliography}

\begin{thebibliography}{55}%
\makeatletter
\providecommand \@ifxundefined [1]{%
 \@ifx{#1\undefined}
}%
\providecommand \@ifnum [1]{%
 \ifnum #1\expandafter \@firstoftwo
 \else \expandafter \@secondoftwo
 \fi
}%
\providecommand \@ifx [1]{%
 \ifx #1\expandafter \@firstoftwo
 \else \expandafter \@secondoftwo
 \fi
}%
\providecommand \natexlab [1]{#1}%
\providecommand \enquote  [1]{``#1''}%
\providecommand \bibnamefont  [1]{#1}%
\providecommand \bibfnamefont [1]{#1}%
\providecommand \citenamefont [1]{#1}%
\providecommand \href@noop [0]{\@secondoftwo}%
\providecommand \href [0]{\begingroup \@sanitize@url \@href}%
\providecommand \@href[1]{\@@startlink{#1}\@@href}%
\providecommand \@@href[1]{\endgroup#1\@@endlink}%
\providecommand \@sanitize@url [0]{\catcode `\\12\catcode `\$12\catcode
  `\&12\catcode `\#12\catcode `\^12\catcode `\_12\catcode `\%12\relax}%
\providecommand \@@startlink[1]{}%
\providecommand \@@endlink[0]{}%
\providecommand \url  [0]{\begingroup\@sanitize@url \@url }%
\providecommand \@url [1]{\endgroup\@href {#1}{\urlprefix }}%
\providecommand \urlprefix  [0]{URL }%
\providecommand \Eprint [0]{\href }%
\providecommand \doibase [0]{http://dx.doi.org/}%
\providecommand \selectlanguage [0]{\@gobble}%
\providecommand \bibinfo  [0]{\@secondoftwo}%
\providecommand \bibfield  [0]{\@secondoftwo}%
\providecommand \translation [1]{[#1]}%
\providecommand \BibitemOpen [0]{}%
\providecommand \bibitemStop [0]{}%
\providecommand \bibitemNoStop [0]{.\EOS\space}%
\providecommand \EOS [0]{\spacefactor3000\relax}%
\providecommand \BibitemShut  [1]{\csname bibitem#1\endcsname}%
\let\auto@bib@innerbib\@empty
\bibitem [{\citenamefont {Elitzur}(1975)}]{Elitzur:1975im}%
  \BibitemOpen
  \bibfield  {author} {\bibinfo {author} {\bibfnamefont {S.}~\bibnamefont
  {Elitzur}},\ }\href {\doibase 10.1103/PhysRevD.12.3978} {\bibfield  {journal}
  {\bibinfo  {journal} {Phys. Rev.}\ }\textbf {\bibinfo {volume} {D12}},\
  \bibinfo {pages} {3978} (\bibinfo {year} {1975})}\BibitemShut {NoStop}%
\bibitem [{\citenamefont {Gribov}(1978)}]{Gribov:1977wm}%
  \BibitemOpen
  \bibfield  {author} {\bibinfo {author} {\bibfnamefont {V.~N.}\ \bibnamefont
  {Gribov}},\ }\href {\doibase 10.1016/0550-3213(78)90175-X} {\bibfield
  {journal} {\bibinfo  {journal} {Nucl. Phys.}\ }\textbf {\bibinfo {volume}
  {B139}},\ \bibinfo {pages} {1} (\bibinfo {year} {1978})},\ \bibinfo {note}
  {[,1(1977)]}\BibitemShut {NoStop}%
\bibitem [{\citenamefont {Singer}(1978)}]{Singer:1978dk}%
  \BibitemOpen
  \bibfield  {author} {\bibinfo {author} {\bibfnamefont {I.~M.}\ \bibnamefont
  {Singer}},\ }\href {\doibase 10.1007/BF01609471} {\bibfield  {journal}
  {\bibinfo  {journal} {Commun. Math. Phys.}\ }\textbf {\bibinfo {volume}
  {60}},\ \bibinfo {pages} {7} (\bibinfo {year} {1978})}\BibitemShut {NoStop}%
\bibitem [{\citenamefont {Fradkin}\ and\ \citenamefont
  {Shenker}(1979)}]{Fradkin:1978dv}%
  \BibitemOpen
  \bibfield  {author} {\bibinfo {author} {\bibfnamefont {E.~H.}\ \bibnamefont
  {Fradkin}}\ and\ \bibinfo {author} {\bibfnamefont {S.~H.}\ \bibnamefont
  {Shenker}},\ }\href {\doibase 10.1103/PhysRevD.19.3682} {\bibfield  {journal}
  {\bibinfo  {journal} {Phys. Rev.}\ }\textbf {\bibinfo {volume} {D19}},\
  \bibinfo {pages} {3682} (\bibinfo {year} {1979})}\BibitemShut {NoStop}%
\bibitem [{\citenamefont {Osterwalder}\ and\ \citenamefont
  {Seiler}(1978)}]{Osterwalder:1977pc}%
  \BibitemOpen
  \bibfield  {author} {\bibinfo {author} {\bibfnamefont {K.}~\bibnamefont
  {Osterwalder}}\ and\ \bibinfo {author} {\bibfnamefont {E.}~\bibnamefont
  {Seiler}},\ }\bibfield  {booktitle} {\emph {\bibinfo {booktitle} {{Salerno
  Soliton Wkshp.1977:0201}}},\ }\href {\doibase 10.1016/0003-4916(78)90039-8}
  {\bibfield  {journal} {\bibinfo  {journal} {Annals Phys.}\ }\textbf {\bibinfo
  {volume} {110}},\ \bibinfo {pages} {440} (\bibinfo {year}
  {1978})}\BibitemShut {NoStop}%
\bibitem [{\citenamefont {Caudy}\ and\ \citenamefont
  {Greensite}(2008)}]{Caudy:2007sf}%
  \BibitemOpen
  \bibfield  {author} {\bibinfo {author} {\bibfnamefont {W.}~\bibnamefont
  {Caudy}}\ and\ \bibinfo {author} {\bibfnamefont {J.}~\bibnamefont
  {Greensite}},\ }\href {\doibase 10.1103/PhysRevD.78.025018} {\bibfield
  {journal} {\bibinfo  {journal} {Phys. Rev.}\ }\textbf {\bibinfo {volume}
  {D78}},\ \bibinfo {pages} {025018} (\bibinfo {year} {2008})},\ \Eprint
  {http://arxiv.org/abs/0712.0999} {arXiv:0712.0999 [hep-lat]} \BibitemShut
  {NoStop}%
\bibitem [{\citenamefont {Friederich}(2013)}]{Friederich:2011xs}%
  \BibitemOpen
  \bibfield  {author} {\bibinfo {author} {\bibfnamefont {S.}~\bibnamefont
  {Friederich}},\ }\href {\doibase 10.1007/s13194-012-0061-y} {\bibfield
  {journal} {\bibinfo  {journal} {Eur. J. Phil. Sci.}\ }\textbf {\bibinfo
  {volume} {3}},\ \bibinfo {pages} {157} (\bibinfo {year} {2013})},\ \Eprint
  {http://arxiv.org/abs/1107.4664} {arXiv:1107.4664 [physics.hist-ph]}
  \BibitemShut {NoStop}%
\bibitem [{\citenamefont {Greensite}\ and\ \citenamefont
  {Matsuyama}(2018)}]{Greensite:2018mhh}%
  \BibitemOpen
  \bibfield  {author} {\bibinfo {author} {\bibfnamefont {J.}~\bibnamefont
  {Greensite}}\ and\ \bibinfo {author} {\bibfnamefont {K.}~\bibnamefont
  {Matsuyama}},\ }\href {\doibase 10.1103/PhysRevD.98.074504} {\bibfield
  {journal} {\bibinfo  {journal} {Phys. Rev.}\ }\textbf {\bibinfo {volume}
  {D98}},\ \bibinfo {pages} {074504} (\bibinfo {year} {2018})},\ \Eprint
  {http://arxiv.org/abs/1805.00985} {arXiv:1805.00985 [hep-th]} \BibitemShut
  {NoStop}%
\bibitem [{\citenamefont {Maas}(2019)}]{Maas:2017wzi}%
  \BibitemOpen
  \bibfield  {author} {\bibinfo {author} {\bibfnamefont {A.}~\bibnamefont
  {Maas}},\ }\href {\doibase 10.1016/j.ppnp.2019.02.003} {\bibfield  {journal}
  {\bibinfo  {journal} {Prog. Part. Nucl. Phys.}\ }\textbf {\bibinfo {volume}
  {106}},\ \bibinfo {pages} {132} (\bibinfo {year} {2019})},\ \Eprint
  {http://arxiv.org/abs/1712.04721} {arXiv:1712.04721 [hep-ph]} \BibitemShut
  {NoStop}%
\bibitem [{\citenamefont {Frohlich}\ \emph {et~al.}(1980)\citenamefont
  {Frohlich}, \citenamefont {Morchio},\ and\ \citenamefont
  {Strocchi}}]{Frohlich:1980gj}%
  \BibitemOpen
  \bibfield  {author} {\bibinfo {author} {\bibfnamefont {J.}~\bibnamefont
  {Frohlich}}, \bibinfo {author} {\bibfnamefont {G.}~\bibnamefont {Morchio}}, \
  and\ \bibinfo {author} {\bibfnamefont {F.}~\bibnamefont {Strocchi}},\ }\href
  {\doibase 10.1016/0370-2693(80)90594-8} {\bibfield  {journal} {\bibinfo
  {journal} {Phys. Lett.}\ }\textbf {\bibinfo {volume} {97B}},\ \bibinfo
  {pages} {249} (\bibinfo {year} {1980})}\BibitemShut {NoStop}%
\bibitem [{\citenamefont {Frohlich}\ \emph {et~al.}(1981)\citenamefont
  {Frohlich}, \citenamefont {Morchio},\ and\ \citenamefont
  {Strocchi}}]{Frohlich:1981yi}%
  \BibitemOpen
  \bibfield  {author} {\bibinfo {author} {\bibfnamefont {J.}~\bibnamefont
  {Frohlich}}, \bibinfo {author} {\bibfnamefont {G.}~\bibnamefont {Morchio}}, \
  and\ \bibinfo {author} {\bibfnamefont {F.}~\bibnamefont {Strocchi}},\ }\href
  {\doibase 10.1016/0550-3213(81)90448-X} {\bibfield  {journal} {\bibinfo
  {journal} {Nucl. Phys.}\ }\textbf {\bibinfo {volume} {B190}},\ \bibinfo
  {pages} {553} (\bibinfo {year} {1981})}\BibitemShut {NoStop}%
\bibitem [{\citenamefont {van Baal}(1992)}]{vanBaal:1991zw}%
  \BibitemOpen
  \bibfield  {author} {\bibinfo {author} {\bibfnamefont {P.}~\bibnamefont {van
  Baal}},\ }\href {\doibase 10.1016/0550-3213(92)90386-P} {\bibfield  {journal}
  {\bibinfo  {journal} {Nucl. Phys.}\ }\textbf {\bibinfo {volume} {B369}},\
  \bibinfo {pages} {259} (\bibinfo {year} {1992})}\BibitemShut {NoStop}%
\bibitem [{\citenamefont {van Baal}(1997)}]{vanBaal:1997gu}%
  \BibitemOpen
  \bibfield  {author} {\bibinfo {author} {\bibfnamefont {P.}~\bibnamefont {van
  Baal}},\ }in\ \href@noop {} {\emph {\bibinfo {booktitle} {{Confinement,
  duality, and nonperturbative aspects of QCD. Proceedings, NATO Advanced Study
  Institute, Newton Institute Workshop, Cambridge, UK, June 23-July 4,
  1997}}}}\ (\bibinfo {year} {1997})\ pp.\ \bibinfo {pages} {161--178},\
  \Eprint {http://arxiv.org/abs/hep-th/9711070} {arXiv:hep-th/9711070 [hep-th]}
  \BibitemShut {NoStop}%
\bibitem [{\citenamefont {Lenz}\ \emph {et~al.}(2000)\citenamefont {Lenz},
  \citenamefont {Negele}, \citenamefont {O'Raifeartaigh},\ and\ \citenamefont
  {Thies}}]{Lenz:2000zt}%
  \BibitemOpen
  \bibfield  {author} {\bibinfo {author} {\bibfnamefont {F.}~\bibnamefont
  {Lenz}}, \bibinfo {author} {\bibfnamefont {J.~W.}\ \bibnamefont {Negele}},
  \bibinfo {author} {\bibfnamefont {L.}~\bibnamefont {O'Raifeartaigh}}, \ and\
  \bibinfo {author} {\bibfnamefont {M.}~\bibnamefont {Thies}},\ }\href
  {\doibase 10.1006/aphy.2000.6072} {\bibfield  {journal} {\bibinfo  {journal}
  {Annals Phys.}\ }\textbf {\bibinfo {volume} {285}},\ \bibinfo {pages} {25}
  (\bibinfo {year} {2000})},\ \Eprint {http://arxiv.org/abs/hep-th/0004200}
  {arXiv:hep-th/0004200 [hep-th]} \BibitemShut {NoStop}%
\bibitem [{\citenamefont {Maas}(2011)}]{Maas:2010nc}%
  \BibitemOpen
  \bibfield  {author} {\bibinfo {author} {\bibfnamefont {A.}~\bibnamefont
  {Maas}},\ }\href {\doibase 10.1140/epjc/s10052-011-1548-y} {\bibfield
  {journal} {\bibinfo  {journal} {Eur. Phys. J.}\ }\textbf {\bibinfo {volume}
  {C71}},\ \bibinfo {pages} {1548} (\bibinfo {year} {2011})},\ \Eprint
  {http://arxiv.org/abs/1007.0729} {arXiv:1007.0729 [hep-lat]} \BibitemShut
  {NoStop}%
\bibitem [{\citenamefont {Capri}\ \emph {et~al.}(2014)\citenamefont {Capri},
  \citenamefont {Dudal}, \citenamefont {Guimaraes}, \citenamefont {Justo},
  \citenamefont {Sorella},\ and\ \citenamefont {Vercauteren}}]{Capri:2013oja}%
  \BibitemOpen
  \bibfield  {author} {\bibinfo {author} {\bibfnamefont {M.~A.~L.}\
  \bibnamefont {Capri}}, \bibinfo {author} {\bibfnamefont {D.}~\bibnamefont
  {Dudal}}, \bibinfo {author} {\bibfnamefont {M.~S.}\ \bibnamefont
  {Guimaraes}}, \bibinfo {author} {\bibfnamefont {I.~F.}\ \bibnamefont
  {Justo}}, \bibinfo {author} {\bibfnamefont {S.~P.}\ \bibnamefont {Sorella}},
  \ and\ \bibinfo {author} {\bibfnamefont {D.}~\bibnamefont {Vercauteren}},\
  }\href {\doibase 10.1016/j.aop.2014.01.014} {\bibfield  {journal} {\bibinfo
  {journal} {Annals Phys.}\ }\textbf {\bibinfo {volume} {343}},\ \bibinfo
  {pages} {72} (\bibinfo {year} {2014})},\ \Eprint
  {http://arxiv.org/abs/1309.1402} {arXiv:1309.1402 [hep-th]} \BibitemShut
  {NoStop}%
\bibitem [{\citenamefont {Sondenheimer}(2020)}]{Sondenheimer:2019idq}%
  \BibitemOpen
  \bibfield  {author} {\bibinfo {author} {\bibfnamefont {R.}~\bibnamefont
  {Sondenheimer}},\ }\href {\doibase 10.1103/PhysRevD.101.056006} {\bibfield
  {journal} {\bibinfo  {journal} {Phys. Rev. D}\ }\textbf {\bibinfo {volume}
  {101}},\ \bibinfo {pages} {056006} (\bibinfo {year} {2020})},\ \Eprint
  {http://arxiv.org/abs/1912.08680} {arXiv:1912.08680 [hep-th]} \BibitemShut
  {NoStop}%
\bibitem [{\citenamefont {Maas}\ \emph
  {et~al.}(2019{\natexlab{a}})\citenamefont {Maas}, \citenamefont
  {Sondenheimer},\ and\ \citenamefont {T\"{o}rek}}]{Maas:2017xzh}%
  \BibitemOpen
  \bibfield  {author} {\bibinfo {author} {\bibfnamefont {A.}~\bibnamefont
  {Maas}}, \bibinfo {author} {\bibfnamefont {R.}~\bibnamefont {Sondenheimer}},
  \ and\ \bibinfo {author} {\bibfnamefont {P.}~\bibnamefont {T\"{o}rek}},\
  }\href {\doibase 10.1016/j.aop.2019.01.010} {\bibfield  {journal} {\bibinfo
  {journal} {Annals Phys.}\ }\textbf {\bibinfo {volume} {402}},\ \bibinfo
  {pages} {18} (\bibinfo {year} {2019}{\natexlab{a}})},\ \Eprint
  {http://arxiv.org/abs/1709.07477} {arXiv:1709.07477 [hep-ph]} \BibitemShut
  {NoStop}%
\bibitem [{\citenamefont {Maas}\ and\ \citenamefont
  {T{\"o}rek}(2017)}]{Maas:2016ngo}%
  \BibitemOpen
  \bibfield  {author} {\bibinfo {author} {\bibfnamefont {A.}~\bibnamefont
  {Maas}}\ and\ \bibinfo {author} {\bibfnamefont {P.}~\bibnamefont
  {T{\"o}rek}},\ }\href {\doibase 10.1103/PhysRevD.95.014501} {\bibfield
  {journal} {\bibinfo  {journal} {Phys. Rev.}\ }\textbf {\bibinfo {volume}
  {D95}},\ \bibinfo {pages} {014501} (\bibinfo {year} {2017})},\ \Eprint
  {http://arxiv.org/abs/1607.05860} {arXiv:1607.05860 [hep-lat]} \BibitemShut
  {NoStop}%
\bibitem [{\citenamefont {T{\"o}rek}\ \emph {et~al.}(2018)\citenamefont
  {T{\"o}rek}, \citenamefont {Maas},\ and\ \citenamefont
  {Sondenheimer}}]{Maas:2017pcw}%
  \BibitemOpen
  \bibfield  {author} {\bibinfo {author} {\bibfnamefont {P.}~\bibnamefont
  {T{\"o}rek}}, \bibinfo {author} {\bibfnamefont {A.}~\bibnamefont {Maas}}, \
  and\ \bibinfo {author} {\bibfnamefont {R.}~\bibnamefont {Sondenheimer}},\
  }\bibfield  {booktitle} {\emph {\bibinfo {booktitle} {{Proceedings, 35th
  International Symposium on Lattice Field Theory (Lattice 2017): Granada,
  Spain, June 18-24, 2017}}},\ }\href {\doibase 10.1051/epjconf/201817508002}
  {\bibfield  {journal} {\bibinfo  {journal} {EPJ Web Conf.}\ }\textbf
  {\bibinfo {volume} {175}},\ \bibinfo {pages} {08002} (\bibinfo {year}
  {2018})},\ \Eprint {http://arxiv.org/abs/1710.01941} {arXiv:1710.01941
  [hep-lat]} \BibitemShut {NoStop}%
\bibitem [{\citenamefont {Maas}\ and\ \citenamefont
  {T\"{o}rek}(2018)}]{Maas:2018xxu}%
  \BibitemOpen
  \bibfield  {author} {\bibinfo {author} {\bibfnamefont {A.}~\bibnamefont
  {Maas}}\ and\ \bibinfo {author} {\bibfnamefont {P.}~\bibnamefont
  {T\"{o}rek}},\ }\href {\doibase 10.1016/j.aop.2018.08.018} {\bibfield
  {journal} {\bibinfo  {journal} {Annals Phys.}\ }\textbf {\bibinfo {volume}
  {397}},\ \bibinfo {pages} {303} (\bibinfo {year} {2018})},\ \Eprint
  {http://arxiv.org/abs/1804.04453} {arXiv:1804.04453 [hep-lat]} \BibitemShut
  {NoStop}%
\bibitem [{\citenamefont {Nielsen}(1975)}]{Nielsen:1975fs}%
  \BibitemOpen
  \bibfield  {author} {\bibinfo {author} {\bibfnamefont {N.~K.}\ \bibnamefont
  {Nielsen}},\ }\href {\doibase 10.1016/0550-3213(75)90301-6} {\bibfield
  {journal} {\bibinfo  {journal} {Nucl. Phys.}\ }\textbf {\bibinfo {volume}
  {B101}},\ \bibinfo {pages} {173} (\bibinfo {year} {1975})}\BibitemShut
  {NoStop}%
\bibitem [{\citenamefont {Gambino}\ and\ \citenamefont
  {Grassi}(2000)}]{Gambino:1999ai}%
  \BibitemOpen
  \bibfield  {author} {\bibinfo {author} {\bibfnamefont {P.}~\bibnamefont
  {Gambino}}\ and\ \bibinfo {author} {\bibfnamefont {P.~A.}\ \bibnamefont
  {Grassi}},\ }\href {\doibase 10.1103/PhysRevD.62.076002} {\bibfield
  {journal} {\bibinfo  {journal} {Phys. Rev. D}\ }\textbf {\bibinfo {volume}
  {62}},\ \bibinfo {pages} {076002} (\bibinfo {year} {2000})},\ \Eprint
  {http://arxiv.org/abs/hep-ph/9907254} {arXiv:hep-ph/9907254} \BibitemShut
  {NoStop}%
\bibitem [{\citenamefont {Dudal}\ \emph {et~al.}(2019)\citenamefont {Dudal},
  \citenamefont {van Egmond}, \citenamefont {Guimares}, \citenamefont
  {Holanda}, \citenamefont {Mintz}, \citenamefont {Palhares}, \citenamefont
  {Peruzzo},\ and\ \citenamefont {Sorella}}]{Dudal:2019aew}%
  \BibitemOpen
  \bibfield  {author} {\bibinfo {author} {\bibfnamefont {D.}~\bibnamefont
  {Dudal}}, \bibinfo {author} {\bibfnamefont {D.~M.}\ \bibnamefont {van
  Egmond}}, \bibinfo {author} {\bibfnamefont {M.~S.}\ \bibnamefont {Guimares}},
  \bibinfo {author} {\bibfnamefont {O.}~\bibnamefont {Holanda}}, \bibinfo
  {author} {\bibfnamefont {B.~W.}\ \bibnamefont {Mintz}}, \bibinfo {author}
  {\bibfnamefont {L.~F.}\ \bibnamefont {Palhares}}, \bibinfo {author}
  {\bibfnamefont {G.}~\bibnamefont {Peruzzo}}, \ and\ \bibinfo {author}
  {\bibfnamefont {S.~P.}\ \bibnamefont {Sorella}},\ }\href {\doibase
  10.1103/PhysRevD.100.065009} {\bibfield  {journal} {\bibinfo  {journal}
  {Phys. Rev.}\ }\textbf {\bibinfo {volume} {D100}},\ \bibinfo {pages} {065009}
  (\bibinfo {year} {2019})},\ \Eprint {http://arxiv.org/abs/1905.10422}
  {arXiv:1905.10422 [hep-th]} \BibitemShut {NoStop}%
\bibitem [{\citenamefont {Dudal}\ \emph
  {et~al.}(2020{\natexlab{a}})\citenamefont {Dudal}, \citenamefont {van
  Egmond}, \citenamefont {Guimaraes}, \citenamefont {Holanda}, \citenamefont
  {Palhares}, \citenamefont {Peruzzo},\ and\ \citenamefont
  {Sorella}}]{Dudal:2019pyg}%
  \BibitemOpen
  \bibfield  {author} {\bibinfo {author} {\bibfnamefont {D.}~\bibnamefont
  {Dudal}}, \bibinfo {author} {\bibfnamefont {D.~M.}\ \bibnamefont {van
  Egmond}}, \bibinfo {author} {\bibfnamefont {M.~S.}\ \bibnamefont
  {Guimaraes}}, \bibinfo {author} {\bibfnamefont {O.}~\bibnamefont {Holanda}},
  \bibinfo {author} {\bibfnamefont {L.~F.}\ \bibnamefont {Palhares}}, \bibinfo
  {author} {\bibfnamefont {G.}~\bibnamefont {Peruzzo}}, \ and\ \bibinfo
  {author} {\bibfnamefont {S.~P.}\ \bibnamefont {Sorella}},\ }\href {\doibase
  10.1007/JHEP02(2020)188} {\bibfield  {journal} {\bibinfo  {journal} {JHEP}\
  }\textbf {\bibinfo {volume} {02}},\ \bibinfo {pages} {188} (\bibinfo {year}
  {2020}{\natexlab{a}})},\ \Eprint {http://arxiv.org/abs/1912.11390}
  {arXiv:1912.11390 [hep-th]} \BibitemShut {NoStop}%
\bibitem [{\citenamefont {Capri}\ \emph {et~al.}(2020)\citenamefont {Capri},
  \citenamefont {Justo}, \citenamefont {Palhares}, \citenamefont {Peruzzo},\
  and\ \citenamefont {Sorella}}]{Capri:2020ppe}%
  \BibitemOpen
  \bibfield  {author} {\bibinfo {author} {\bibfnamefont {M.}~\bibnamefont
  {Capri}}, \bibinfo {author} {\bibfnamefont {I.}~\bibnamefont {Justo}},
  \bibinfo {author} {\bibfnamefont {L.}~\bibnamefont {Palhares}}, \bibinfo
  {author} {\bibfnamefont {G.}~\bibnamefont {Peruzzo}}, \ and\ \bibinfo
  {author} {\bibfnamefont {S.}~\bibnamefont {Sorella}},\ }\href {\doibase
  10.1103/PhysRevD.102.033003} {\bibfield  {journal} {\bibinfo  {journal}
  {Phys. Rev. D}\ }\textbf {\bibinfo {volume} {102}},\ \bibinfo {pages}
  {033003} (\bibinfo {year} {2020})},\ \Eprint
  {http://arxiv.org/abs/2007.01770} {arXiv:2007.01770 [hep-th]} \BibitemShut
  {NoStop}%
\bibitem [{\citenamefont {Denner}(1993)}]{Denner:1991kt}%
  \BibitemOpen
  \bibfield  {author} {\bibinfo {author} {\bibfnamefont {A.}~\bibnamefont
  {Denner}},\ }\href {\doibase 10.1002/prop.2190410402} {\bibfield  {journal}
  {\bibinfo  {journal} {Fortsch. Phys.}\ }\textbf {\bibinfo {volume} {41}},\
  \bibinfo {pages} {307} (\bibinfo {year} {1993})},\ \Eprint
  {http://arxiv.org/abs/0709.1075} {arXiv:0709.1075 [hep-ph]} \BibitemShut
  {NoStop}%
\bibitem [{\citenamefont {Sirlin}(1991)}]{Sirlin:1991rt}%
  \BibitemOpen
  \bibfield  {author} {\bibinfo {author} {\bibfnamefont {A.}~\bibnamefont
  {Sirlin}},\ }\href {\doibase 10.1016/0370-2693(91)91254-S} {\bibfield
  {journal} {\bibinfo  {journal} {Phys. Lett. B}\ }\textbf {\bibinfo {volume}
  {267}},\ \bibinfo {pages} {240} (\bibinfo {year} {1991})}\BibitemShut
  {NoStop}%
\bibitem [{\citenamefont {Stuart}(1991{\natexlab{a}})}]{Stuart:1991xk}%
  \BibitemOpen
  \bibfield  {author} {\bibinfo {author} {\bibfnamefont {R.~G.}\ \bibnamefont
  {Stuart}},\ }\href {\doibase 10.1016/0370-2693(91)90653-8} {\bibfield
  {journal} {\bibinfo  {journal} {Phys. Lett. B}\ }\textbf {\bibinfo {volume}
  {262}},\ \bibinfo {pages} {113} (\bibinfo {year}
  {1991}{\natexlab{a}})}\BibitemShut {NoStop}%
\bibitem [{\citenamefont {Stuart}(1991{\natexlab{b}})}]{Stuart:1991cc}%
  \BibitemOpen
  \bibfield  {author} {\bibinfo {author} {\bibfnamefont {R.~G.}\ \bibnamefont
  {Stuart}},\ }\href {\doibase 10.1016/0370-2693(91)91842-J} {\bibfield
  {journal} {\bibinfo  {journal} {Phys. Lett. B}\ }\textbf {\bibinfo {volume}
  {272}},\ \bibinfo {pages} {353} (\bibinfo {year}
  {1991}{\natexlab{b}})}\BibitemShut {NoStop}%
\bibitem [{\citenamefont {Aeppli}\ \emph {et~al.}(1994)\citenamefont {Aeppli},
  \citenamefont {van Oldenborgh},\ and\ \citenamefont {Wyler}}]{Aeppli:1993rs}%
  \BibitemOpen
  \bibfield  {author} {\bibinfo {author} {\bibfnamefont {A.}~\bibnamefont
  {Aeppli}}, \bibinfo {author} {\bibfnamefont {G.~J.}\ \bibnamefont {van
  Oldenborgh}}, \ and\ \bibinfo {author} {\bibfnamefont {D.}~\bibnamefont
  {Wyler}},\ }\href {\doibase 10.1016/0550-3213(94)90195-3} {\bibfield
  {journal} {\bibinfo  {journal} {Nucl. Phys. B}\ }\textbf {\bibinfo {volume}
  {428}},\ \bibinfo {pages} {126} (\bibinfo {year} {1994})},\ \Eprint
  {http://arxiv.org/abs/hep-ph/9312212} {arXiv:hep-ph/9312212} \BibitemShut
  {NoStop}%
\bibitem [{\citenamefont {Kniehl}\ and\ \citenamefont
  {Sirlin}(1998)}]{Kniehl:1998fn}%
  \BibitemOpen
  \bibfield  {author} {\bibinfo {author} {\bibfnamefont {B.~A.}\ \bibnamefont
  {Kniehl}}\ and\ \bibinfo {author} {\bibfnamefont {A.}~\bibnamefont
  {Sirlin}},\ }\href {\doibase 10.1103/PhysRevLett.81.1373} {\bibfield
  {journal} {\bibinfo  {journal} {Phys. Rev. Lett.}\ }\textbf {\bibinfo
  {volume} {81}},\ \bibinfo {pages} {1373} (\bibinfo {year} {1998})},\ \Eprint
  {http://arxiv.org/abs/hep-ph/9805390} {arXiv:hep-ph/9805390} \BibitemShut
  {NoStop}%
\bibitem [{\citenamefont {Denner}\ \emph {et~al.}(2005)\citenamefont {Denner},
  \citenamefont {Dittmaier}, \citenamefont {Roth},\ and\ \citenamefont
  {Wieders}}]{Denner:2005fg}%
  \BibitemOpen
  \bibfield  {author} {\bibinfo {author} {\bibfnamefont {A.}~\bibnamefont
  {Denner}}, \bibinfo {author} {\bibfnamefont {S.}~\bibnamefont {Dittmaier}},
  \bibinfo {author} {\bibfnamefont {M.}~\bibnamefont {Roth}}, \ and\ \bibinfo
  {author} {\bibfnamefont {L.}~\bibnamefont {Wieders}},\ }\href {\doibase
  10.1016/j.nuclphysb.2011.09.001} {\bibfield  {journal} {\bibinfo  {journal}
  {Nucl. Phys. B}\ }\textbf {\bibinfo {volume} {724}},\ \bibinfo {pages} {247}
  (\bibinfo {year} {2005})},\ \bibinfo {note} {[Erratum: Nucl.Phys.B 854,
  504--507 (2012)]},\ \Eprint {http://arxiv.org/abs/hep-ph/0505042}
  {arXiv:hep-ph/0505042} \BibitemShut {NoStop}%
\bibitem [{\citenamefont {Denner}\ and\ \citenamefont
  {Dittmaier}(2006)}]{Denner:2006ic}%
  \BibitemOpen
  \bibfield  {author} {\bibinfo {author} {\bibfnamefont {A.}~\bibnamefont
  {Denner}}\ and\ \bibinfo {author} {\bibfnamefont {S.}~\bibnamefont
  {Dittmaier}},\ }\href {\doibase 10.1016/j.nuclphysbps.2006.09.025} {\bibfield
   {journal} {\bibinfo  {journal} {Nucl. Phys. B Proc. Suppl.}\ }\textbf
  {\bibinfo {volume} {160}},\ \bibinfo {pages} {22} (\bibinfo {year} {2006})},\
  \Eprint {http://arxiv.org/abs/hep-ph/0605312} {arXiv:hep-ph/0605312}
  \BibitemShut {NoStop}%
\bibitem [{\citenamefont {Bredenstein}\ \emph {et~al.}(2006)\citenamefont
  {Bredenstein}, \citenamefont {Denner}, \citenamefont {Dittmaier},\ and\
  \citenamefont {Weber}}]{Bredenstein:2006rh}%
  \BibitemOpen
  \bibfield  {author} {\bibinfo {author} {\bibfnamefont {A.}~\bibnamefont
  {Bredenstein}}, \bibinfo {author} {\bibfnamefont {A.}~\bibnamefont {Denner}},
  \bibinfo {author} {\bibfnamefont {S.}~\bibnamefont {Dittmaier}}, \ and\
  \bibinfo {author} {\bibfnamefont {M.}~\bibnamefont {Weber}},\ }\href
  {\doibase 10.1103/PhysRevD.74.013004} {\bibfield  {journal} {\bibinfo
  {journal} {Phys. Rev. D}\ }\textbf {\bibinfo {volume} {74}},\ \bibinfo
  {pages} {013004} (\bibinfo {year} {2006})},\ \Eprint
  {http://arxiv.org/abs/hep-ph/0604011} {arXiv:hep-ph/0604011} \BibitemShut
  {NoStop}%
\bibitem [{\citenamefont {Kniehl}\ and\ \citenamefont
  {Sirlin}(2008)}]{Kniehl:2008cj}%
  \BibitemOpen
  \bibfield  {author} {\bibinfo {author} {\bibfnamefont {B.~A.}\ \bibnamefont
  {Kniehl}}\ and\ \bibinfo {author} {\bibfnamefont {A.}~\bibnamefont
  {Sirlin}},\ }\href {\doibase 10.1103/PhysRevD.77.116012} {\bibfield
  {journal} {\bibinfo  {journal} {Phys. Rev. D}\ }\textbf {\bibinfo {volume}
  {77}},\ \bibinfo {pages} {116012} (\bibinfo {year} {2008})},\ \Eprint
  {http://arxiv.org/abs/0801.0669} {arXiv:0801.0669 [hep-th]} \BibitemShut
  {NoStop}%
\bibitem [{\citenamefont {Denner}\ and\ \citenamefont
  {Lang}(2015)}]{Denner:2014zga}%
  \BibitemOpen
  \bibfield  {author} {\bibinfo {author} {\bibfnamefont {A.}~\bibnamefont
  {Denner}}\ and\ \bibinfo {author} {\bibfnamefont {J.-N.}\ \bibnamefont
  {Lang}},\ }\href {\doibase 10.1140/epjc/s10052-015-3579-2} {\bibfield
  {journal} {\bibinfo  {journal} {Eur. Phys. J. C}\ }\textbf {\bibinfo {volume}
  {75}},\ \bibinfo {pages} {377} (\bibinfo {year} {2015})},\ \Eprint
  {http://arxiv.org/abs/1406.6280} {arXiv:1406.6280 [hep-ph]} \BibitemShut
  {NoStop}%
\bibitem [{\citenamefont {Grassi}\ \emph {et~al.}(2002)\citenamefont {Grassi},
  \citenamefont {Kniehl},\ and\ \citenamefont {Sirlin}}]{Grassi:2001bz}%
  \BibitemOpen
  \bibfield  {author} {\bibinfo {author} {\bibfnamefont {P.~A.}\ \bibnamefont
  {Grassi}}, \bibinfo {author} {\bibfnamefont {B.~A.}\ \bibnamefont {Kniehl}},
  \ and\ \bibinfo {author} {\bibfnamefont {A.}~\bibnamefont {Sirlin}},\ }\href
  {\doibase 10.1103/PhysRevD.65.085001} {\bibfield  {journal} {\bibinfo
  {journal} {Phys. Rev. D}\ }\textbf {\bibinfo {volume} {65}},\ \bibinfo
  {pages} {085001} (\bibinfo {year} {2002})},\ \Eprint
  {http://arxiv.org/abs/hep-ph/0109228} {arXiv:hep-ph/0109228} \BibitemShut
  {NoStop}%
\bibitem [{\citenamefont {Haussling}\ and\ \citenamefont
  {Kraus}(1997)}]{Haussling:1996rq}%
  \BibitemOpen
  \bibfield  {author} {\bibinfo {author} {\bibfnamefont {R.}~\bibnamefont
  {Haussling}}\ and\ \bibinfo {author} {\bibfnamefont {E.}~\bibnamefont
  {Kraus}},\ }\href {\doibase 10.1007/s002880050521} {\bibfield  {journal}
  {\bibinfo  {journal} {Z. Phys. C}\ }\textbf {\bibinfo {volume} {75}},\
  \bibinfo {pages} {739} (\bibinfo {year} {1997})},\ \Eprint
  {http://arxiv.org/abs/hep-th/9608160} {arXiv:hep-th/9608160} \BibitemShut
  {NoStop}%
\bibitem [{\citenamefont {Dudal}\ \emph
  {et~al.}(2020{\natexlab{b}})\citenamefont {Dudal}, \citenamefont {van
  Egmond}, \citenamefont {Guimaraes}, \citenamefont {Palhares}, \citenamefont
  {Peruzzo},\ and\ \citenamefont {Sorella}}]{Dudal:2020uwb}%
  \BibitemOpen
  \bibfield  {author} {\bibinfo {author} {\bibfnamefont {D.}~\bibnamefont
  {Dudal}}, \bibinfo {author} {\bibfnamefont {D.}~\bibnamefont {van Egmond}},
  \bibinfo {author} {\bibfnamefont {M.}~\bibnamefont {Guimaraes}}, \bibinfo
  {author} {\bibfnamefont {L.}~\bibnamefont {Palhares}}, \bibinfo {author}
  {\bibfnamefont {G.}~\bibnamefont {Peruzzo}}, \ and\ \bibinfo {author}
  {\bibfnamefont {S.}~\bibnamefont {Sorella}},\ }\href@noop {} {\  (\bibinfo
  {year} {2020}{\natexlab{b}})},\ \Eprint {http://arxiv.org/abs/2008.07813}
  {arXiv:2008.07813 [hep-th]} \BibitemShut {NoStop}%
\bibitem [{\citenamefont {Papavassiliou}\ and\ \citenamefont
  {Pilaftsis}(1995)}]{Papavassiliou:1995fq}%
  \BibitemOpen
  \bibfield  {author} {\bibinfo {author} {\bibfnamefont {J.}~\bibnamefont
  {Papavassiliou}}\ and\ \bibinfo {author} {\bibfnamefont {A.}~\bibnamefont
  {Pilaftsis}},\ }\href {\doibase 10.1103/PhysRevLett.75.3060} {\bibfield
  {journal} {\bibinfo  {journal} {Phys. Rev. Lett.}\ }\textbf {\bibinfo
  {volume} {75}},\ \bibinfo {pages} {3060} (\bibinfo {year} {1995})},\ \Eprint
  {http://arxiv.org/abs/hep-ph/9506417} {arXiv:hep-ph/9506417} \BibitemShut
  {NoStop}%
\bibitem [{\citenamefont {Papavassiliou}\ and\ \citenamefont
  {Pilaftsis}(1996{\natexlab{a}})}]{Papavassiliou:1995gs}%
  \BibitemOpen
  \bibfield  {author} {\bibinfo {author} {\bibfnamefont {J.}~\bibnamefont
  {Papavassiliou}}\ and\ \bibinfo {author} {\bibfnamefont {A.}~\bibnamefont
  {Pilaftsis}},\ }\href {\doibase 10.1103/PhysRevD.53.2128} {\bibfield
  {journal} {\bibinfo  {journal} {Phys. Rev. D}\ }\textbf {\bibinfo {volume}
  {53}},\ \bibinfo {pages} {2128} (\bibinfo {year} {1996}{\natexlab{a}})},\
  \Eprint {http://arxiv.org/abs/hep-ph/9507246} {arXiv:hep-ph/9507246}
  \BibitemShut {NoStop}%
\bibitem [{\citenamefont {Papavassiliou}\ and\ \citenamefont
  {Pilaftsis}(1996{\natexlab{b}})}]{Papavassiliou:1996zn}%
  \BibitemOpen
  \bibfield  {author} {\bibinfo {author} {\bibfnamefont {J.}~\bibnamefont
  {Papavassiliou}}\ and\ \bibinfo {author} {\bibfnamefont {A.}~\bibnamefont
  {Pilaftsis}},\ }\href {\doibase 10.1103/PhysRevD.54.5315} {\bibfield
  {journal} {\bibinfo  {journal} {Phys. Rev. D}\ }\textbf {\bibinfo {volume}
  {54}},\ \bibinfo {pages} {5315} (\bibinfo {year} {1996}{\natexlab{b}})},\
  \Eprint {http://arxiv.org/abs/hep-ph/9605385} {arXiv:hep-ph/9605385}
  \BibitemShut {NoStop}%
\bibitem [{\citenamefont {Papavassiliou}\ and\ \citenamefont
  {Pilaftsis}(1998)}]{Papavassiliou:1997pb}%
  \BibitemOpen
  \bibfield  {author} {\bibinfo {author} {\bibfnamefont {J.}~\bibnamefont
  {Papavassiliou}}\ and\ \bibinfo {author} {\bibfnamefont {A.}~\bibnamefont
  {Pilaftsis}},\ }\href {\doibase 10.1103/PhysRevD.58.053002} {\bibfield
  {journal} {\bibinfo  {journal} {Phys. Rev. D}\ }\textbf {\bibinfo {volume}
  {58}},\ \bibinfo {pages} {053002} (\bibinfo {year} {1998})},\ \Eprint
  {http://arxiv.org/abs/hep-ph/9710426} {arXiv:hep-ph/9710426} \BibitemShut
  {NoStop}%
\bibitem [{\citenamefont {Binosi}\ and\ \citenamefont
  {Papavassiliou}(2009)}]{Binosi:2009qm}%
  \BibitemOpen
  \bibfield  {author} {\bibinfo {author} {\bibfnamefont {D.}~\bibnamefont
  {Binosi}}\ and\ \bibinfo {author} {\bibfnamefont {J.}~\bibnamefont
  {Papavassiliou}},\ }\href {\doibase 10.1016/j.physrep.2009.05.001} {\bibfield
   {journal} {\bibinfo  {journal} {Phys. Rept.}\ }\textbf {\bibinfo {volume}
  {479}},\ \bibinfo {pages} {1} (\bibinfo {year} {2009})},\ \Eprint
  {http://arxiv.org/abs/0909.2536} {arXiv:0909.2536 [hep-ph]} \BibitemShut
  {NoStop}%
\bibitem [{\citenamefont {Wurtz}\ and\ \citenamefont
  {Lewis}(2013)}]{Wurtz:2013ova}%
  \BibitemOpen
  \bibfield  {author} {\bibinfo {author} {\bibfnamefont {M.}~\bibnamefont
  {Wurtz}}\ and\ \bibinfo {author} {\bibfnamefont {R.}~\bibnamefont {Lewis}},\
  }\href {\doibase 10.1103/PhysRevD.88.054510} {\bibfield  {journal} {\bibinfo
  {journal} {Phys. Rev. D}\ }\textbf {\bibinfo {volume} {88}},\ \bibinfo
  {pages} {054510} (\bibinfo {year} {2013})},\ \Eprint
  {http://arxiv.org/abs/1307.1492} {arXiv:1307.1492 [hep-lat]} \BibitemShut
  {NoStop}%
\bibitem [{\citenamefont {Maas}\ and\ \citenamefont
  {Mufti}(2015)}]{Maas:2014pba}%
  \BibitemOpen
  \bibfield  {author} {\bibinfo {author} {\bibfnamefont {A.}~\bibnamefont
  {Maas}}\ and\ \bibinfo {author} {\bibfnamefont {T.}~\bibnamefont {Mufti}},\
  }\href {\doibase 10.1103/PhysRevD.91.113011} {\bibfield  {journal} {\bibinfo
  {journal} {Phys. Rev.}\ }\textbf {\bibinfo {volume} {D91}},\ \bibinfo {pages}
  {113011} (\bibinfo {year} {2015})},\ \Eprint {http://arxiv.org/abs/1412.6440}
  {arXiv:1412.6440 [hep-lat]} \BibitemShut {NoStop}%
\bibitem [{\citenamefont {Maas}(2013)}]{Maas:2012tj}%
  \BibitemOpen
  \bibfield  {author} {\bibinfo {author} {\bibfnamefont {A.}~\bibnamefont
  {Maas}},\ }\href {\doibase 10.1142/S0217732313501034} {\bibfield  {journal}
  {\bibinfo  {journal} {Mod. Phys. Lett.}\ }\textbf {\bibinfo {volume} {A28}},\
  \bibinfo {pages} {1350103} (\bibinfo {year} {2013})},\ \Eprint
  {http://arxiv.org/abs/1205.6625} {arXiv:1205.6625 [hep-lat]} \BibitemShut
  {NoStop}%
\bibitem [{\citenamefont {Maas}\ and\ \citenamefont
  {Mufti}(2014)}]{Maas:2013aia}%
  \BibitemOpen
  \bibfield  {author} {\bibinfo {author} {\bibfnamefont {A.}~\bibnamefont
  {Maas}}\ and\ \bibinfo {author} {\bibfnamefont {T.}~\bibnamefont {Mufti}},\
  }\href {\doibase 10.1007/JHEP04(2014)006} {\bibfield  {journal} {\bibinfo
  {journal} {JHEP}\ }\textbf {\bibinfo {volume} {04}},\ \bibinfo {pages} {006}
  (\bibinfo {year} {2014})},\ \Eprint {http://arxiv.org/abs/1312.4873}
  {arXiv:1312.4873 [hep-lat]} \BibitemShut {NoStop}%
\bibitem [{\citenamefont {Maas}\ \emph
  {et~al.}(2019{\natexlab{b}})\citenamefont {Maas}, \citenamefont {Raubitzek},\
  and\ \citenamefont {T\"orek}}]{Maas:2018ska}%
  \BibitemOpen
  \bibfield  {author} {\bibinfo {author} {\bibfnamefont {A.}~\bibnamefont
  {Maas}}, \bibinfo {author} {\bibfnamefont {S.}~\bibnamefont {Raubitzek}}, \
  and\ \bibinfo {author} {\bibfnamefont {P.}~\bibnamefont {T\"orek}},\ }\href
  {\doibase 10.1103/PhysRevD.99.074509} {\bibfield  {journal} {\bibinfo
  {journal} {Phys. Rev. D}\ }\textbf {\bibinfo {volume} {99}},\ \bibinfo
  {pages} {074509} (\bibinfo {year} {2019}{\natexlab{b}})},\ \Eprint
  {http://arxiv.org/abs/1811.03395} {arXiv:1811.03395 [hep-lat]} \BibitemShut
  {NoStop}%
\bibitem [{\citenamefont {Maas}(2017)}]{Maas:2017csm}%
  \BibitemOpen
  \bibfield  {author} {\bibinfo {author} {\bibfnamefont {A.}~\bibnamefont
  {Maas}},\ }\href {\doibase 10.1016/j.aop.2017.10.003} {\bibfield  {journal}
  {\bibinfo  {journal} {Annals Phys.}\ }\textbf {\bibinfo {volume} {387}},\
  \bibinfo {pages} {29} (\bibinfo {year} {2017})},\ \Eprint
  {http://arxiv.org/abs/1705.03812} {arXiv:1705.03812 [hep-lat]} \BibitemShut
  {NoStop}%
\bibitem [{\citenamefont {Heinzl}\ \emph {et~al.}(2008)\citenamefont {Heinzl},
  \citenamefont {Ilderton}, \citenamefont {Langfeld}, \citenamefont {Lavelle},\
  and\ \citenamefont {McMullan}}]{Heinzl:2008bu}%
  \BibitemOpen
  \bibfield  {author} {\bibinfo {author} {\bibfnamefont {T.}~\bibnamefont
  {Heinzl}}, \bibinfo {author} {\bibfnamefont {A.}~\bibnamefont {Ilderton}},
  \bibinfo {author} {\bibfnamefont {K.}~\bibnamefont {Langfeld}}, \bibinfo
  {author} {\bibfnamefont {M.}~\bibnamefont {Lavelle}}, \ and\ \bibinfo
  {author} {\bibfnamefont {D.}~\bibnamefont {McMullan}},\ }\href {\doibase
  10.1103/PhysRevD.78.074511} {\bibfield  {journal} {\bibinfo  {journal} {Phys.
  Rev.}\ }\textbf {\bibinfo {volume} {D78}},\ \bibinfo {pages} {074511}
  (\bibinfo {year} {2008})},\ \Eprint {http://arxiv.org/abs/0807.4698}
  {arXiv:0807.4698 [hep-lat]} \BibitemShut {NoStop}%
\bibitem [{\citenamefont {Maas}(2016)}]{Maas:2016edk}%
  \BibitemOpen
  \bibfield  {author} {\bibinfo {author} {\bibfnamefont {A.}~\bibnamefont
  {Maas}},\ }\href {\doibase 10.1140/epjc/s10052-016-4216-4} {\bibfield
  {journal} {\bibinfo  {journal} {Eur. Phys. J. C}\ }\textbf {\bibinfo {volume}
  {76}},\ \bibinfo {pages} {366} (\bibinfo {year} {2016})},\ \Eprint
  {http://arxiv.org/abs/1603.07525} {arXiv:1603.07525 [hep-lat]} \BibitemShut
  {NoStop}%
\bibitem [{\citenamefont {Egger}\ \emph {et~al.}(2017)\citenamefont {Egger},
  \citenamefont {Maas},\ and\ \citenamefont {Sondenheimer}}]{Egger:2017tkd}%
  \BibitemOpen
  \bibfield  {author} {\bibinfo {author} {\bibfnamefont {L.}~\bibnamefont
  {Egger}}, \bibinfo {author} {\bibfnamefont {A.}~\bibnamefont {Maas}}, \ and\
  \bibinfo {author} {\bibfnamefont {R.}~\bibnamefont {Sondenheimer}},\ }\href
  {\doibase 10.1142/S0217732317502121} {\bibfield  {journal} {\bibinfo
  {journal} {Mod. Phys. Lett.}\ }\textbf {\bibinfo {volume} {A32}},\ \bibinfo
  {pages} {1750212} (\bibinfo {year} {2017})},\ \Eprint
  {http://arxiv.org/abs/1701.02881} {arXiv:1701.02881 [hep-ph]} \BibitemShut
  {NoStop}%
\bibitem [{\citenamefont {Fernbach}\ \emph {et~al.}(2020)\citenamefont
  {Fernbach}, \citenamefont {Lechner}, \citenamefont {Maas}, \citenamefont
  {Pl{\"a}tzer},\ and\ \citenamefont {Sch{\"o}fbeck}}]{Fernbach:2020tpa}%
  \BibitemOpen
  \bibfield  {author} {\bibinfo {author} {\bibfnamefont {S.}~\bibnamefont
  {Fernbach}}, \bibinfo {author} {\bibfnamefont {L.}~\bibnamefont {Lechner}},
  \bibinfo {author} {\bibfnamefont {A.}~\bibnamefont {Maas}}, \bibinfo {author}
  {\bibfnamefont {S.}~\bibnamefont {Pl{\"a}tzer}}, \ and\ \bibinfo {author}
  {\bibfnamefont {R.}~\bibnamefont {Sch{\"o}fbeck}},\ }\href {\doibase
  10.1103/PhysRevD.101.114018} {\bibfield  {journal} {\bibinfo  {journal}
  {Phys. Rev. D}\ }\textbf {\bibinfo {volume} {101}},\ \bibinfo {pages}
  {114018} (\bibinfo {year} {2020})},\ \Eprint
  {http://arxiv.org/abs/2002.01688} {arXiv:2002.01688 [hep-ph]} \BibitemShut
  {NoStop}%
\end{thebibliography}%

\end{document}